\documentclass[%
reprint,
superscriptaddress,
nofootinbib,
nobibnotes,
 amsmath,amssymb,
 aps,
 prd,
 longbibliography,
 twocolumn,
]{revtex4-2}

\usepackage{float}
\usepackage[percent]{overpic}
\usepackage{graphicx, comment}
\usepackage{dcolumn}
\usepackage{bm}
\usepackage{xcolor}
\usepackage{hyperref}
\usepackage[mathlines]{lineno}
\usepackage[normalem]{ulem}
\usepackage{amsfonts}
\usepackage{url}
\usepackage{enumitem}

\begin{document}

\newlength{\figwidth}
\newlength{\fighalfwidth}
\setlength{\figwidth}{0.95\textwidth}
\setlength{\fighalfwidth}{0.5\textwidth}
\newcommand{\blue}[1]{{\textcolor{blue}{#1}}}
\newcommand{\red}[1]{{\textcolor{red}{#1}}}
\newcommand{\fixme}[1]{\textit{\textcolor{red}{Fixme: #1}}}

\newcommand{\Ar}{\mathrm{Ar}}
\newcommand{\Eavail}{E_\mathrm{avail}}
\newcommand{\Edep}{E_\mathrm{dep}}
\newcommand{\Evis}{E_\mathrm{vis}}
\newcommand{\Erec}{E_\mathrm{rec}}
\newcommand{\deltaCP}{\delta_{\mathrm{CP}}}
\newcommand{\firstOM}{$1^{\mathrm{st}}$~OM}
\newcommand{\secondOM}{$2^{\mathrm{nd}}$~OM}


\title{Self-compensating Light Calorimetry with Liquid Argon Time Projection Chamber for GeV Neutrino Physics}

\newcommand{\BNL}{Brookhaven National Laboratory, Upton, New York 11973, USA}
\newcommand{\SBU}{Stony Brook University, SUNY, Stony Brook, New York, 11794, USA}
\author{Xuyang Ning} \affiliation{\BNL}
\author{Wei Shi} \affiliation{\SBU}
\author{Chao Zhang} \email{czhang@bnl.gov} \affiliation{\BNL} 
\author{Ciro Riccio} \affiliation{\SBU}
\author{Jay Hyun Jo} \affiliation{\BNL}


\date{\today}
\begin{abstract}
The Liquid Argon Time Projection Chamber (LArTPC) is a powerful dual calorimeter capable of estimating particle energy from both ionization charge and scintillation light. 
Our study shows that, due to the recombination luminescence, the LArTPC functions as a self-compensating light calorimeter: the missing energy in the hadronic component is compensated for by the increased luminescence relative to the electromagnetic component. 
Using 0.5--5~GeV electron neutrino charged current interactions as a case study, we show that good compensation of the electron-to-hadron response ratio ($e/h$) from 1--1.05 can be achieved across a broad range of drift electric fields (0.2--1.8~kV/cm), with better performance for neutrino energies above 2~GeV.
This study highlights the potential of light calorimetry in LArTPCs for GeV neutrino energy reconstruction, complementing traditional charge calorimetry.
Under ideal conditions of uniform light collection, we show that LArTPC light calorimetry can achieve an energy resolution comparable to the charge imaging calorimetry. Challenges arising from nonuniform light collection in large LArTPCs can be mitigated with a position-dependent light yield correction derived from 3D charge signal imaging.
\end{abstract}

\maketitle


\section{Introduction}
\label{sec:intro}

Estimating the energy of incident particles is an essential task in high energy physics experiments to study the fundamental properties of particles. For example, in neutrino oscillation experiments, the oscillation probabilities are expressed as a function of distance over neutrino energy ($L/E_\nu$). Therefore, accurately estimating the neutrino energy with minimum bias and high resolution is a goal of all neutrino detectors. 
Calorimetry is an important detection principle in particle physics to measure particle energy~\cite{Fabjan:2003aq}. 
A particle loses its energy in the calorimeter primarily through electromagnetic or hadronic interactions. The energy deposited by the charged particles in the instrumented material of the calorimeter can be detected in measurable signals, such as charge, light, heat, and so on. These signals are then used to estimate the initial energy of the incident particles after calibration and correcting for the missing energy not measured in the calorimeter.

Liquid argon (LAr) was proposed for calorimetry in 1974 by Willis and Radeka~\cite{willis74} and since then has been used by many particle physics experiments~\cite{Bonivento:2024qpn}, such as E706~\cite{E706:1997rip}, SLD~\cite{Axen:1990dz}, D0~\cite{D0:1992ftr}, H1~\cite{H1CalorimeterGroup:1993boq}, and most recently, ATLAS~\cite{ATLAS:2010blk}. 
As a noble liquid detector, LAr produces large signals of both ionization charge and scintillation light when energy is deposited inside, although almost all previous collider LAr calorimeters are ionization chambers that only collect ionization electrons with short drift distances. 
Because of the relatively long radiation length of LAr and the size constraint in collider experiments, those LAr ionization chambers also consist of denser absorbers (e.g.~copper or tungsten) to contain the energy and act as sampling calorimeters. 
This LAr sampling calorimeter solution is cost-effective, provides good energy resolution, linearity, uniformity, fast response, and has negligible radiation damage. Therefore, it is particularly attractive to high energy (GeV to TeV) and high luminosity accelerator experiments.

The technology of using LAr as a scintillation calorimeter has been mainly driven by dark matter experiments, such as DEAP-3600~\cite{DEAP:2019yzn}, WArP~\cite{Acciarri:2010zz}, DarkSide-50~\cite{DarkSide:2014llq}, and DarkSide-20k~\cite{DarkSide-20k:2017zyg}. Since dark matter experiments desire low energy thresholds (tens of keV) and the capability of distinguishing electronic recoils from nuclear recoils, much effort was invested in understanding the LAr scintillation mechanism, light yield, scintillation timing, response to energy transfer, optical transport properties, and so on. 
LAr scintillation calorimeter for few-hundred-MeV or higher particle energy measurement has not been fully studied or utilized so far compared to other noble liquid detectors such as LKr~\cite{NA48:2007mvn} or LXe~\cite{MEG:2016leq}, mainly because of its longer radiation length resulting larger sized detector needed. 

In recent years, the technology to build massive and homogeneous liquid argon time projection chambers (LArTPCs) has advanced tremendously, owing to its many advantages in neutrino detection such as the cost-effectiveness, scalability, and high spatial resolution for particle identification. The concept was first proposed in the 1970s~\cite{rubbia77, Chen:1976pp} and the first large LArTPC detector, ICARUS T600~\cite{ICARUS:2004wqc}, operated in 2010. 
Many LArTPC experiments have been built since then, such as MicroBooNE~\cite{MicroBooNE:2016pwy}, ICARUS~\cite{ICARUS-T600:2020ajz}, SBND~\cite{SBND:2020scp}, ProtoDUNE-SP~\cite{DUNE:2021hwx}, and the upcoming Deep Underground Neutrino Experiment (DUNE)~\cite{DUNE:2020ypp}. 
In the current-generation LArTPC neutrino experiments, the ionization charge signal is the main detection channel. With a uniform electric field applied in the TPC between the cathode and anode planes, the ionization electrons drift toward the anode at a constant speed. 
The drift time of the electron can be translated into its position along the drift direction. The transverse positions can be determined by the multiple wire planes, or printed circuit boards, at the anode where the parallel sense wires or strips are oriented at different angles. 
Pixelated electrodes at the anode have also been developed in recent years~\cite{DUNE:2024fjn}. Together, a 3D image of the particle activities can be reconstructed with a millimeter-scale position resolution. 
Meanwhile, the energy deposition per unit length ($dE/dx$), also called the linear energy transfer (LET), can be measured along the particle trajectories to provide further particle identification information. 
Since high purity can be achieved in LAr, electrons can drift a few meters without attenuation, making it possible to build a large, fully active LArTPC calorimeter for neutrino detection.

Scintillation light is used in LArTPC neutrino detectors primarily to provide the starting time ($t_0$) of the drift since the ionization charge signal does not provide the $t_0$ information by itself. Light calorimetry was not fully utilized or studied in previous LArTPC neutrino detectors primarily because of their low photon collection efficiency and large position nonuniformity. 
This situation may be improved in future LArTPC detectors. 
For instance, the proposed reference design of the DUNE Phase-II far detectors, namely, the Aluminum Profiles with Embedded X-Arapucas (APEX) design~\cite{DUNE:2024wvj}, aims to increase the optical surface coverage to about 60\% by instrumenting the entire field cage walls with large area photon detectors. 
Preliminary simulations show that the average light yield of this design can reach about 180 photoelectrons per MeV, similar in performance to other organic liquid scintillator detectors used in neutrino experiments. Therefore, it is timely to revisit the LArTPC as a scintillation light calorimeter and study how well it compares to and supplements the well-established ionization charge calorimetry in LArTPC.  

This paper is organized as follows. Section~\ref{sec:energy_dissipation} briefly reviews the energy dissipation mechanism in LAr. In Sec.~\ref{sec:energy-reco}, the deposited energy and missing energy in LArTPC is estimated for GeV neutrinos as a case study using Monte Carlo simulations. Section~\ref{sec:L_calo} studies in detail the self-compensation mechanism observed in the LArTPC light calorimetry. Section~\ref{sec:light-calo-performance} compares the energy resolution achievable using light calorimetry with charge-based calorimetry in reconstructing neutrino energy. 
Finally, Sec.~\ref{sec:conclusions} discusses the applications of light calorimetry in the current and future experiments.

\section{Energy dissipation in liquid argon}
\label{sec:energy_dissipation}

The fundamental properties of liquid argon as a particle radiation detector have been extensively studied since the 1970s. In this section, we briefly review the energy dissipation mechanisms in LAr to establish the groundwork for later sections. More details can be found in Refs.~\cite{Doke:1981eac,Doke:1990rza,Segreto:2020qks,Szydagis5010013}.
Table~\ref{tab:symbols} lists the symbols used in this paper, including their values, definitions, and descriptions.

\begin{table}[htpb]
\centering
\begin{tabular}{c|c|l}
\hline\hline
   Symbol  & Definition & Description \\
   \hline
    $\alpha$ & 0.21 & $N_{ex} / N_i$  \\
    $\beta$ & 0.83 & $1/(1+\alpha) $\\
    $W_q$ & 19.5~eV & energy to create one quantum \\
    $W_\mathrm{ion}$ & 23.6~eV & $W_q/\beta$ \\
    $N_i$ & Eq.~\ref{eq:N_eg} & number of $e^-$--$\Ar^+$ pairs  \\
    $N_{ex}$ & Eq.~\ref{eq:N_eg} & number of $\Ar^*$  \\
    $N_q$ & $N_{i}+N_{ex}$ & total number of quanta  \\
    \hline
    $R_c$ & Eq.~\ref{eq:Rc} & charge recombination factor \\
    $R_L$ & Eq.~\ref{eq:R_L} & light recombination factor \\
    $N_e$ & $N_i R_c$ & electrons after recombination \\
    $N_{ph}$ & $N_q R_L$ & photons after recombination  \\
    \hline
    $E_\mathrm{avail}$ &  & total available energy \\
    $E_\mathrm{dep}$ &  & deposited energy in LAr  \\
    $R_\mathrm{dep}$ &$E_\mathrm{dep}/E_\mathrm{avail}$ & fraction of energy not missing \\
    $E_\mathrm{vis}$ & & visible (measured) energy \\
    $Q$ & Eq.~\ref{eq:QL} & visible energy in charge   \\
    $L$ & Eq.~\ref{eq:QL} & visible energy in light   \\
    $E_\mathrm{rec}$ & & reconstructed energy \\
    \hline
    $R_\mathrm{cal}$ &$E_\mathrm{vis}/E_\mathrm{avail}$ & calorimeter response \\
    $e$ & & EM component \\
    $h$ & & hadronic component \\
    $e/h$ & $R_\mathrm{cal}^e/R_\mathrm{cal}^h$ & compensation performance metric \\
    $f_{em}$ & & fraction of $E_\mathrm{vis}$ in $e$ \\
    \hline
    $\epsilon$ & & photon collection efficiency (PCE) \\
    $N_{pe}$ & $\epsilon N_{ph}$ & number of detected photoelectrons  \\

\hline\hline
\end{tabular}
\caption{List of symbols used in this paper. The symbols are organized into sections based on their relevance to quanta creation, recombination, energy reconstruction, calorimeter response, and others.}
\label{tab:symbols}
\end{table}

Energy deposition in LAr creates either excited argon atoms (Ar$^*$) or electron-ion (e$^-$--Ar$^+$) pairs. 
The ratio $\alpha$ between the number of Ar$^*$, $N_{ex}$, and the number of Ar$^+$, $N_i$, is calculated and measured to be 0.21~\cite{Doke:1990rza}:
\begin{equation}
    \alpha = \frac{N_{ex}}{N_i} = 0.21.
\end{equation}
The energy required to produce one electron-ion pair in LAr is measured to be $W_\mathrm{ion} = 23.6$~eV~\cite{TAKAHASHI1973123}. If we treat the Ar$^*$ and Ar$^+$ equally as one quantized state, then it follows that the average energy, or a $quantum$, required to produce one state is:
\begin{equation}
    W_\mathrm{q} = \frac{W_\mathrm{ion}}{(1+\alpha)} = \beta \cdot W_\mathrm{ion} = 19.5~\mathrm{eV},
\end{equation}
where $\beta = 1/(1+\alpha) = 0.83$. 
Since $W_q$ also corresponds to the ideal energy needed to create one photon at zero electric field (cf.~Eq.~\ref{eq:R_L}), it is called $W_{ph}$ in some literatures such as Ref.~\cite{Doke:1990rza}.  

\begin{figure}[htp]
  \centering
    \includegraphics[width=0.9\columnwidth]{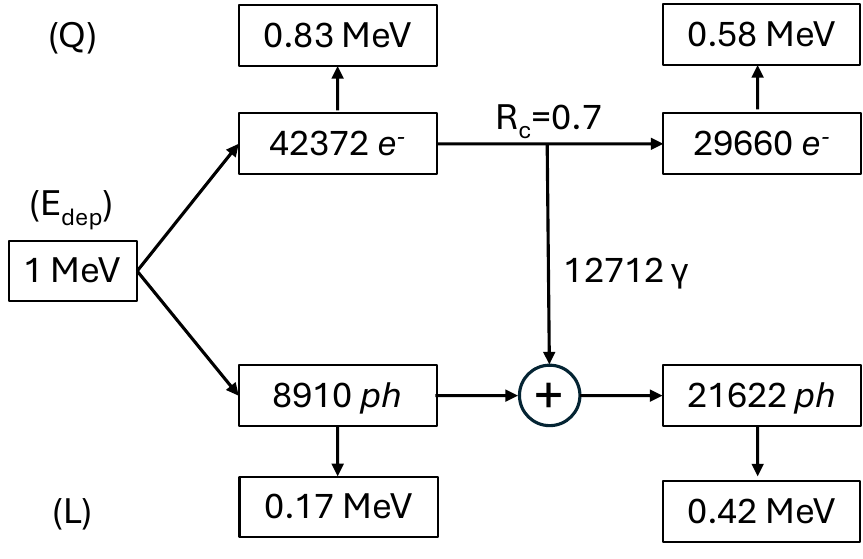}
  \caption{Illustration of the distribution of quanta between ionization electrons ($e^-$) and scintillation photons ($ph$) before and after charge recombination in LAr. The energy deposition ($\Edep$) is assumed to be 1~MeV and the charge recombination factor is assumed to be $R_c=0.7$. The number of quanta is converted into the visible energy in charge ($Q$) and light ($L$) assuming each quantum equaling $W_q=19.5$~eV.
  }
  \label{fig:edep_lar}
\end{figure}

The ionization electrons in the electron-ion pairs can be collected under an external drift electric field and form the signals in charge calorimetry. The Ar$^*$ will de-excite through the process~\cite{Doke:2002oab}: 
\begin{eqnarray}    \label{eq:de-ex}
    \mathrm{Ar}^* + \mathrm{Ar} & \to & \mathrm{Ar}_2^* \\ \nonumber
    \mathrm{Ar}_2^* & \to & 2\mathrm{Ar} + ph,
\end{eqnarray}
where a scintillation photon ($ph$) is emitted in the process. The wavelength of the emission spectrum peaks at 127~nm, in the vacuum ultraviolet region. The collection of these scintillation photons forms the signals in light calorimetry. 
The number of ionization electrons and scintillation photons created by a certain energy deposition ($\Edep$) in LAr can be calculated as:
\begin{eqnarray} \label{eq:N_eg}
    N_i & = & \beta \cdot \frac{\Edep}{W_q} \\ \nonumber
    N_{ex} & = & (1-\beta) \cdot \frac{\Edep}{W_q} \\ \nonumber
    N_q & = & N_i + N_{ex} = \frac{\Edep}{W_q}, 
\end{eqnarray}
where $N_q$ is the total number of quanta, including both the ionization electrons and the scintillation photons. For 1~MeV deposited energy, $N_q$ is about 51000, which consists of $\sim$42000 electrons and $\sim$9000 photons. This initial distribution of quanta is illustrated in Fig.~\ref{fig:edep_lar}.

The ionization electrons and argon ions, however, go through another process called recombination during their passage in LAr~\cite{Doke:2002oab}:
\begin{eqnarray}
    \Ar^+ + \Ar & \to &  \Ar_2^+ \\ \nonumber
    \Ar_2^+ + e^- & \to &  \Ar^{**} + \Ar  \\ \nonumber
     \Ar^{**} & \to &  \Ar^{*} + \mathrm{heat}.
\end{eqnarray}
The resulting $\Ar^*$ goes through the same de-excitation process described in Eq.~\ref{eq:de-ex} and releases one scintillation photon. Therefore, the recombination process consumes the ionization electrons but creates more scintillation photons. 
The increase of scintillation light from recombination, also called recombination luminescence, was first observed in 1978 by S. Kubota \textit{et al.}~\cite{Kubota:1978oha}, and the recombination mechanism was well articulated by A.~Hitachi \textit{et al.} in 1992~\cite{Hitachi:1992iyr} and further elaborated by T.~Doke \textit{et al.} in 2002~\cite{Doke:2002oab}. The redistribution of quanta after recombination is illustrated in Fig.~\ref{fig:edep_lar}. It is worth noting that the total number of quanta, $N_q$, is conserved before and after the recombination process. The conservation of quanta in LAr has been observed experimentally at a range of LETs and electric fields for various particles~\cite{Doke:2002oab}.

The mechanism of charge recombination in noble liquid has been studied in many literatures~\cite{Thomas:1987zz,Segreto:2024xnp}. The charge recombination factor $R_c$, defined as the ratio between the number of ionization electrons after recombination ($N_e$) and before recombination ($N_i$), is well described by the Birks model~\cite{Thomas:1987zz}:
\begin{equation} \label{eq:Rc}
    R_c = \frac{N_e}{N_i} = 0.8 / \left(1 + \frac{0.0486}{\mathcal{E}\rho} \cdot \frac{dE}{dx} \right),
\end{equation}
where $\mathcal{E}$ is the external electric field in kV/cm, $\rho$ is the density of LAr in g/cm$^{3}$, and $dE/dx$ is the energy loss per unit length in MeV/cm. For a minimum ionizing particle (MIP), the mean $dE/dx$ in LAr is about 2.1~MeV/cm, and the recombination factor is about 0.7 at a typical 0.5~kV/cm electric field in LArTPC neutrino experiments. 

\begin{figure}[htpb]
  \centering
    \includegraphics[width=\columnwidth]{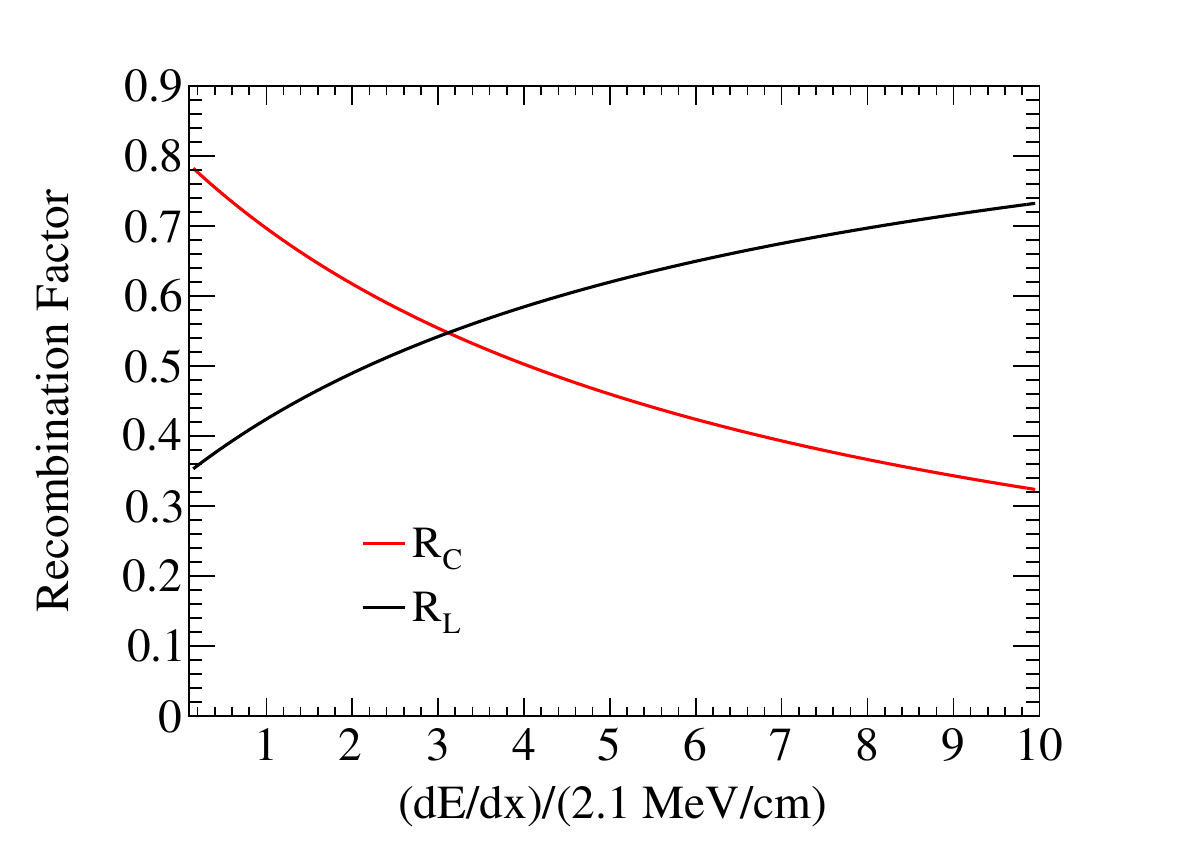}
  \caption{Recombination factor for charge ($R_c$) and light ($R_L$) as a function of $dE/dx$ in LArTPC based on Eqs.~\ref{eq:Rc} and \ref{eq:R_L}. The external drift electric field is assumed to be 0.5~kV/cm. The $dE/dx$ is normalized to that of a minimum ionizing particle, which is about 2.1~MeV/cm.
  }
  \label{fig:recob}
\end{figure}

At zero electric field, Eq.~\ref{eq:Rc} indicates that all ionization electrons will recombine, and an equal amount of scintillation photons will be generated. The ideal photon yield is therefore expected to be $N_q$, i.e.~about 51000 photons per MeV of deposited energy at zero electric field\footnote{The effects to reduce the photon yield at zero E-field, as well as other quanta reducing processes, are discussed in Appendix \ref{sec:app_quanta_reducing}.}. From this we can derive the light recombination factor $R_L$, defined as the ratio between the number of photons after recombination ($N_{ph}$) and the ideal photon yield, to be:
\begin{equation} \label{eq:R_L}
    R_L = \frac{N_{ph}}{N_q} = 1 - \beta R_c.
\end{equation}
Figure \ref{fig:recob} shows the charge and light recombination factors as a function of $dE/dx$ in LArTPC, assuming a drift electric field of 0.5~kV/cm, where the $dE/dx$ is normalized to that of a MIP. The recombination effect is the opposite between the charge and the light: The percentage of ionization electrons after recombination decreases for heavily ionizing particles compared with a MIP, while the percentage of scintillation photons increases. 
The scintillation light yields in LAr for some light hadrons compared to electrons have been measured experimentally in the 1970s--1980s and summarized by T.~Doke \textit{et al.}~\cite{Doke:2002oab}. These theoretical and experimental results form the basis of the self-compensation mechanism in LArTPC light calorimetry, which will be further described in Sec.~\ref{sec:L_calo}.

\section{Energy estimation in LArTPC calorimeters}
\label{sec:energy-reco}

The goal of the energy reconstruction is to estimate the initial energy of the incident particles from the observed signals in the calorimeter. A LArTPC can be considered as a dual calorimeter because both the charge and the light signals are recorded and can be used for energy estimation. 

Energy reconstruction is typically performed in two steps. First, the energy deposited in the active material of the calorimeter is estimated from the observed signals in charge or light. 
Then, the initial energy of the incident particles is estimated by correcting for contribution from the invisible energy, or missing energy, which corresponds to the energy loss that does not create a signal in the calorimeter. 
Missing energy can originate from several mechanisms, such as energy loss in the inactive materials, energy spent to break up the nuclei, escaping neutral particles such as neutrinos from particle decays, low-energy particles below the detection thresholds, and so on. Correcting missing energy generally requires a comprehensive detector Monte Carlo simulation study. 
In this section, we describe the methods involved in estimating the deposited energy and the missing energy in LArTPC.

\subsection{Deposited energy estimation}
\label{sec:edep_estimation}

A LArTPC records charge and light signals as digitized waveforms. The charge sensors record induced electrical signals when ionization electrons pass by them. After dedicated signal processing and charge calibration~\cite{MicroBooNE:2018swd}, the signal waveforms can be converted to the number of ionization electrons detected, which is the same as $N_e$ if we neglect the detection threshold. The light sensors record the photoelectron-induced signals when optical photons strike a sensor's photosensitive surfaces. Similarly, after signal processing and single photoelectron calibration, the signal waveforms can be converted to the number of photoelectrons ($N_{pe}$) detected, which can be further converted into the total number of optical photons ($N_{ph}$) given a calibrated photon collection efficiency (PCE) $\epsilon$: $N_{ph} = N_{pe}/\epsilon$.

There are different conventions to convert $N_e$ and $N_{ph}$ to the visible energy in charge ($Q$) and light ($L$), which only differ numerically before $Q$ and $L$ are further converted to the deposited energy $\Edep$. 
To take advantage of the conservation of quanta as explained in Sec.~\ref{sec:energy_dissipation}, we recommend that each quantum, whether of an ionization electron or a scintillation photon, represents the same amount of energy. 
The most convenient choice of that energy is $W_q$, or 19.5~eV. With this definition, $Q$ and $L$ become:
\begin{equation} \label{eq:QL}
    Q = W_q \cdot N_e, \quad
    L = W_q \cdot N_{ph}.
\end{equation}

Both $Q$ and $L$ can be used to estimate the deposited energy by correcting for the recombination effect described in Sec.~\ref{sec:energy_dissipation}:
\begin{equation}  \label{eq:E_dep}
  E _\mathrm{dep} = \frac{Q}{\beta R_c} 
  = \frac{L}{1 - \beta R_c}.
\end{equation}
Since the charge recombination factor $R_c$ is dependent on the $dE/dx$ (cf.~Eq.~\ref{eq:Rc}), without further corrections at the particle trajectory level, the $\Edep$ estimation with an average $R_c$ suffers from event-by-event fluctuations. The smearing of $\Edep$ from charge recombination can be removed if we sum $Q$ and $L$ together:
\begin{eqnarray}  \label{eq:E_dep_QL}
  Q + L & = & W_q \cdot (N_e + N_{ph}) \\ \nonumber 
  & = & W_q \cdot (N_i + N_{ex}) \\ \nonumber
  & = & E _\mathrm{dep}.
\end{eqnarray}
In this case, the conservation of quanta is transferred to the conservation of $Q+L$, and the $\Edep$ estimation is free of the recombination effect. Figure \ref{fig:edep_lar} illustrates the estimation of $Q$ and $L$ for a 1~MeV deposited energy with $R_c=0.7$, and $Q+L$ is conserved before and after recombination.

Eq.~\ref{eq:E_dep} and Eq.~\ref{eq:E_dep_QL} show that there are three ways to estimate the $\Edep$: charge-only ($Q$), light-only ($L$), or charge-plus-light ($Q+L$). It may appear that $Q+L$, being free of the recombination effect, is the best energy estimator. However, since the final goal is to reconstruct the initial energy of the incident particles, we have to consider the contribution from missing energy, which often is the dominant factor in energy resolution.

\subsection{Missing energy estimation}
\label{sec:missing_energy}

It is expected that some energy loss of the particles does not produce detectable signals and is, therefore, missing. 
Estimating this missing energy is essential in energy reconstruction and it is typically done through a detailed detector simulation using software packages such as GEANT4~\cite{GEANT4:2002zbu}. There are several contributors to the missing energy in general particle calorimeters:

\paragraph{Inactive material} 
In sampling calorimeters, dense absorber materials are purposely placed in the detector to degrade the particle energy and contain the electromagnetic and hadronic showers. Absorber and active layers are typically alternately placed and only the active layers are instrumented to record the energy deposited inside. 
LArTPC, in contrast, can be built as a homogeneously active calorimeter. In this study, we consider an ideal, infinitely large LArTPC, and no inactive material is involved. In reality, a detector has a finite size and could consist of multiple modulized LArTPCs, so the energy loss in the structural materials, as well as the partially contained events, needs to be studied. It is also worth noting that the active volume for charge and light calorimetry could be different since scintillator light can be generated in the LAr region where there is no electric field, which is not considered in this study.

\paragraph{Detection threshold} As particles slow down, their kinetic energy eventually becomes too small to generate a detectable signal. For charge calorimetry, the detection threshold of individual energy deposition depends on the electronic noise level. In most recent LArTPCs, such as MicroBooNE~\cite{MicroBooNE:2017qiu} and ProtoDUNE-SP~\cite{DUNE:2020cqd}, the equivalent noise charge has achieved below 400 $e^-$ owing to the implementation of cryogenic electronics. In this study, we set the charge detection threshold at 3000 $e^-$ for any energy deposition within 5~mm. The corresponding energy threshold is particle dependent, but it is approximately equal to $\Edep=100$~keV for a MIP particle with  $R_c=0.7$. This charge threshold setting is inspired by the low energy studies in recent LArTPC experiments~\cite{Castiglioni:2020tsu}. For light calorimetry, we assume each photon detector's detection threshold is much below one photoelectron, so there is no loss of light signal below the threshold for GeV events.

\paragraph{Nuclear breakup} Since argon has a composite nucleus, particles could inelastically scatter on Ar and break up the nucleus to produce protons, neutrons, pions, and so on. These secondary hadrons tend to interact with Ar and produce more hadrons, which result in a hadronic shower. While doing so, some energy of the particles is spent to overcome the nuclear binding energy and becomes part of the missing energy. Particularly, the fraction of energy spent on the invisible nuclear breakup could be 30\%--50\% for neutrons~\cite{Friedland:2018vry}, since they are neutral particles and don't deposit energy directly. Moreover, the event-by-event fluctuation for nuclear breakup is large and often becomes the dominant factor in the resolution of energy reconstruction. 

\paragraph{Particle decay and others} There are several other mechanisms to generate missing energy. The most common one is the charged pion and muon decays where some energy is carried away by neutrinos that escape the detector. Negatively charged pions and muons can also be captured by the Ar nuclei, resulting in some missing energy when the nuclei de-excite~\cite{LArIAT:2024otd}. Neutrons will capture on Ar after thermalization and release several $\gamma$ rays that add up to about 6~MeV, which become the ``negative'' missing energy, i.e.~extra energy if the captured $\gamma$ rays are included in the calorimetry.

In the following, we use GeV electron neutrinos as a case study to estimate missing energy contribution and energy responses in LArTPC. Muon neutrinos are also studied with similar conclusions and not reported here. Energetic neutrinos from 0.5--5~GeV are sent into a LArTPC detector. This energy range is consistent with the wide-band neutrino beam planned with the DUNE experiment~\cite{DUNE:2020ypp} and $\nu_e$ is the main appearance signal in DUNE. The LAr volume is large enough so that all particle activities are contained. The $v$-Ar interaction is modeled by the GENIE neutrino event generator~\cite{Andreopoulos:2009rq} version 2.12.10. Only the charged current (CC) interaction is simulated since this is the main detection channel in neutrino oscillation studies for flavor tagging. 
The propagation of the primary particles in LAr after the $\nu_e$-Ar CC interaction is performed through the EDEP-SIM software package~\cite{edep-sim}, which provides a convenient interface to GEANT4. GEANT4 version 4.10.3.p01b is used in this study. The energy depositions per 5~mm step (or less if the final step is shorter than 5~mm) along the particle trajectories are stored. The $\Edep$ in each step is then converted to the charge and light signals in terms of the number of ionization electrons and scintillation photons, taking into account the recombination effect, based on the procedures described in Sec.~\ref{sec:energy_dissipation}. The drift electric field is set to be 0.5~kV/cm to calculate the charge recombination factor $R_c$. 

The number of scintillation photons after recombination ($N_{ph}$) is further converted to the total number of detected photoelectrons ($N_{pe}$) assuming an average absolute PCE of 0.8\%. 
This PCE is based on a standalone GEANT4 simulation study of the APEX design for DUNE Phase-II far detectors~\cite{DUNE:2024wvj}, where the surface optical coverage is $\sim$60\% and the photon detector (X-Arapuca) efficiency is $\sim$2\%. 
In the current APEX design, the PCE is nonuniform across the detector volume because of the specific placements of the photon detectors. 
However, in this study, we assume a uniform PCE as an ideal case. 
Under this PCE, the light yield is about 180 photoelectrons per MeV of deposited energy for a MIP. Methods to mitigate nonuniform light collection are discussed in Sec.~\ref{sec:app_nonuniformity}. Event-by-event fluctuation is applied to $N_{pe}$ assuming a Poisson distribution, although it does not contribute significantly to the energy resolution for GeV neutrino events at this light yield. Finally, the simulated $N_e$ and $N_{pe}$ are converted to the visible energy in charge ($Q$) and light ($L$) following Eq.~\ref{eq:QL}.

\begin{figure}[htp]
  \centering
        \includegraphics[width=0.9\columnwidth]{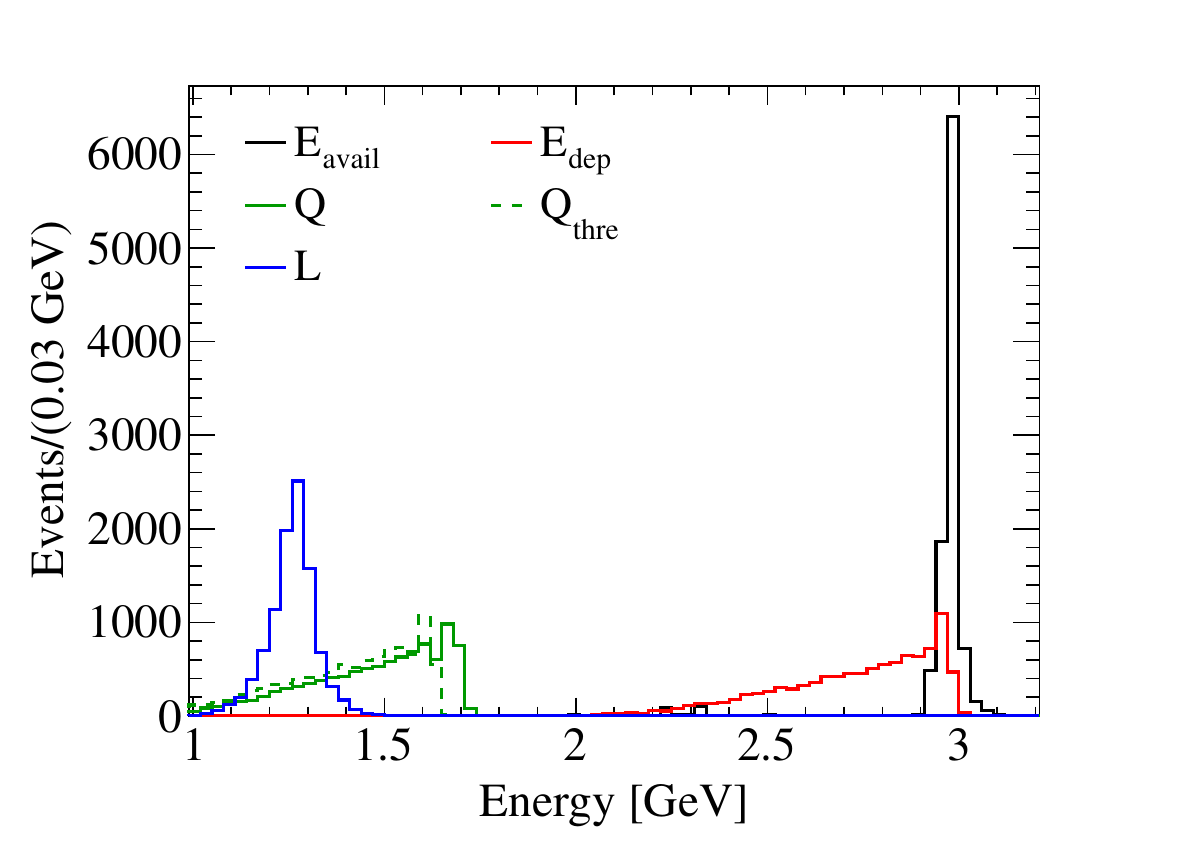}
  \caption{Simulated distribution of energy dissipation from $10^4$ 3~GeV $\nu_e$-Ar CC interactions. Black histogram shows the distribution of the total available energy ($\Eavail$) of final state particles from the GENIE event generator. Red histogram shows the distribution of the deposited energy ($\Edep$) in LAr. The green and blue histograms show the distributions of visible energy in charge ($Q$) and light ($L$), respectively, after the charge recombination. The dotted green histogram shows the distribution of $Q$ after applying the charge detection threshold of 3000 ionization electrons.
  }
  \label{fig:edep_QL}
\end{figure}

Figure.~\ref{fig:edep_QL} shows the simulated results for a 3~GeV electron neutrino, which is close to the peak energy of the DUNE neutrino beam. The $\nu_e$-Ar CC interaction can be generally written as: 
\begin{equation} \label{eq:nueCC}
    \nu_e + Ar \to e^- + \pi^0\mathrm{s} + \mathrm{hadrons} + X,
\end{equation}
where hadrons could include protons, neutrons, charged pions, and so on, and $X$ is the nucleus after the interaction. Zero or more $\pi^0$s can be produced in the interaction and together with the electron from the $\nu_e$CC interaction, they form the electromagnetic activities in the event. 
In the GeV neutrino energy range, the $\nu_e$-Ar interaction cross section consists of several components such as the quasielastic scattering, the $\Delta$ resonance scattering, the deep inelastic scattering, and so on. These interactions are modeled by the GENIE event generator to produce the final state particles and their kinematic distributions. The black histogram in Fig.~\ref{fig:edep_QL} shows the distribution of the total available energy ($E_{\mathrm{avail}}$) of the final state particles after the $\nu_e$-Ar interaction. $E_{\mathrm{avail}}$ includes the kinematic energy of stable or long-lived particles such as the protons, neutrons, and electrons, and the total energy of short-lived particles including their masses such as pions and muons. Some energy is missing at the $\nu_e$-Ar interaction due to nuclear breakup, but the energy smearing is relatively small at the neutrino generator level.

The red histogram in Fig.~\ref{fig:edep_QL} shows the distribution of the deposited energy ($\Edep$) in LAr. The missing energy is obvious at this step and the event-by-event fluctuation is large. The distribution has a long tail extending to lower energy, which is mostly caused by nuclear breakup when the primary or secondary hadrons, especially neutrons, inelastically scatter with Ar. This long low-energy tail creates an undesired feed-down effect in physics analysis since a high-energy neutrino event could appear at a much lower deposited energy. 

\begin{figure}[htp]
  \centering
    \includegraphics[width=0.9\columnwidth]{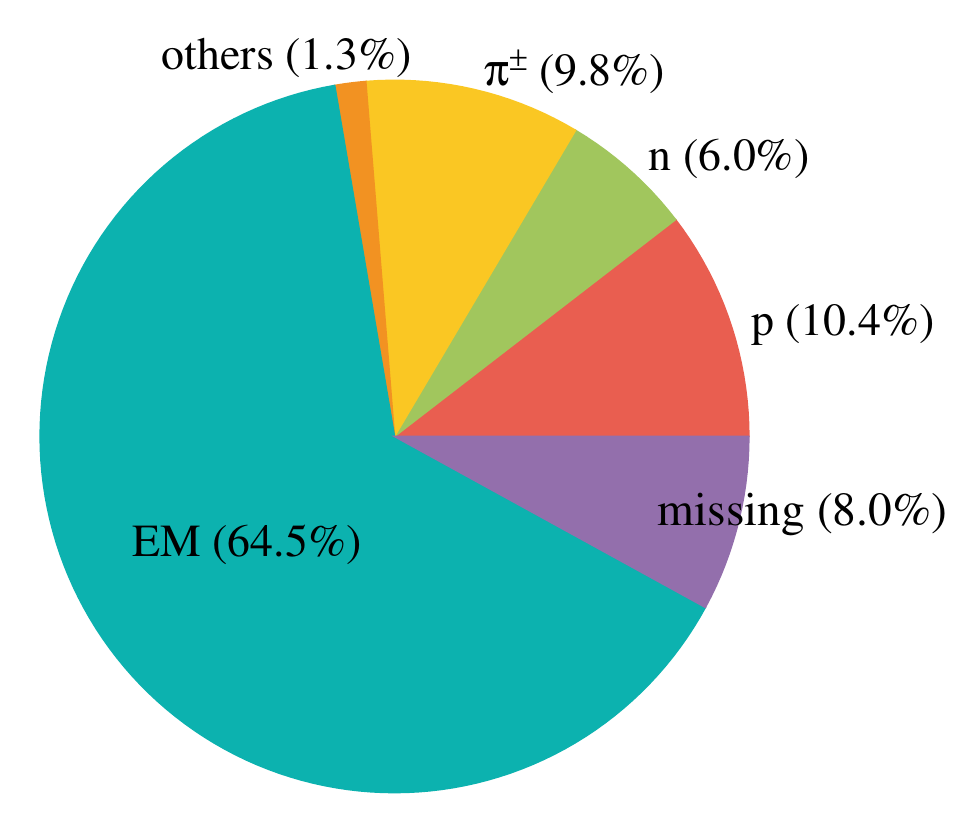}
    \includegraphics[width=0.9\columnwidth]{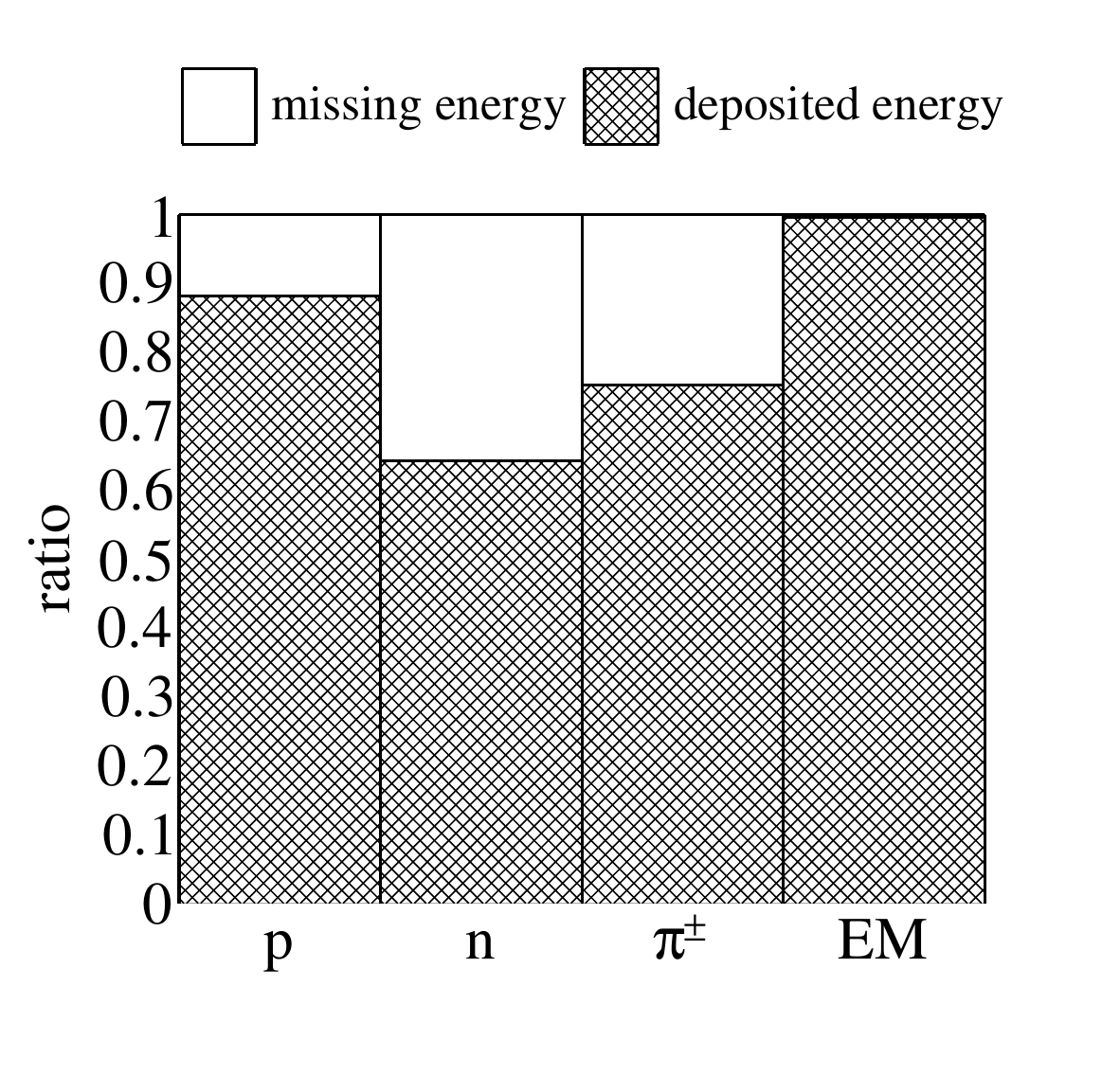}
  \caption{(Top) Energy distribution in a $\nu_e$-Ar CC interaction. The pie chart shows how the total available energy ($E_\mathrm{avail}$) in a $\nu_e$-Ar CC interaction is divided, based on an average of $10^4$ simulated 3 GeV $\nu_e$ CC events (cf. Eq.~\ref{eq:nueCC}). The $E_\mathrm{avail}$ is allocated across various categories of primary particles' deposited energy (protons, neutrons, charged pions, EM components, and others) as well as the total missing energy. For each primary particle, its deposited energy includes contributions from both the particle itself and its secondary particles as they interact with LAr.
  (Bottom) For each category of primary particles, their available energy is further divided into deposited energy and missing energy. The respective energy ratios are displayed for each particle category.
  }
  \label{fig:E-component}
\end{figure}

Figure \ref{fig:E-component} further illustrates the missing energy contribution in $\nu_e$-Ar CC interactions. 
The pie chart in the top panel shows how the total available energy ($E_\mathrm{avail}$) in a $\nu_e$-Ar CC interaction is divided, based on an average of $10^4$ simulated 3 GeV $\nu_e$ CC events (cf. Eq.~\ref{eq:nueCC}). 
The $E_\mathrm{avail}$ is allocated across various categories of primary particles and the total missing energy.
For each primary particle, its deposited energy includes contributions from both the particle itself and its secondary particles as they interact with LAr.
On average, 64.5\% of $\Eavail$ is distributed in the EM component, consisting of electrons and $\pi^0$s. The hadronic component is divided into protons, neutrons, charged pions, and other particles such as kaons and lambda baryons, where their deposited energy over $\Eavail$ are 10.4\%, 6.0\%, 9.8\%, and 1.3\%, respectively. The total missing energy that is unmeasured is 8.0\% on average. In the bottom panel, the available energy of each primary particle after a $\nu_e$-Ar CC interaction is divided into its deposited energy and missing energy. The average percentage of missing energy is 11.8\%, 35.8\%, 24.7\%, and 0.04\% for protons, neutrons, charged pions, and the EM component, respectively, indicating that missing energy mostly comes from the hadronic component.

The green and blue histograms in Fig.~\ref{fig:edep_QL} show the visible energy measured in $Q$ and $L$, respectively, with the simulation steps described earlier that include the recombination effect. 
The $Q$ distribution (green) is relatively broader than the $\Edep$ distribution (red), which is caused by the additional smearing from the $dE/dx$ dependent charge recombination. 
The $Q$ distribution after applying the charge detection threshold (3000 $e^-$) is shown as the dotted green histogram, and the missing energy at this detection threshold is about 3\%. 
Interestingly, the $L$ distribution (blue) does not have the low-energy tail, is more symmetric, and has a much smaller event-by-event fluctuation than $Q$. The origin of this desired feature in LArTPC light calorimetry  will be explored in Sec.~\ref{sec:L_calo}.

Using the simulation results such as shown in Fig.~\ref{fig:edep_QL}, we can correct for the missing energy bias by applying an additional factor to scale the mean of each distribution ($Q$, $L$, or $\Edep$) to the initial neutrino energy $E_\nu$. The event-by-event fluctuation, however, cannot be corrected for and it contributes to the overall resolution of the energy reconstruction. 
The energy resolution is better using light calorimetry ($L$) than the simple charge calorimetry ($Q$), which will be further elaborated in Sec.~\ref{sec:L_calo}. 
It is worth pointing out that the $Q+L$ method using Eq.~\ref{eq:E_dep_QL}, while removing the recombination effect and recovering the $\Edep$ distribution, does not significantly improve upon the charge-only calorimetry $Q$, because the missing energy contribution dominates the energy resolution. There are, however, other methods to further improve energy resolution in charge calorimetry by taking advantage of the imaging capability of LArTPC, which is not the main topic of this article and is described in Appendix \ref{sec:app_charge_imaging}.

\section{Self-compensating light calorimetry}
\label{sec:L_calo}

The features of the LArTPC light calorimetry seen in Fig.~\ref{fig:edep_QL} can be understood in analogous to the \emph{compensating calorimeter} concept in HEP accelerator experiments, which is an important technique because of the fundamental problems of hadron calorimetry~\cite{Fabjan:2003aq,Wigmans:2018fua}. 
The development of hadronic cascades results in both the electromagnetic (EM) component and the hadronic component. The EM component ($e$) comes from the $\pi^0$ and $\eta$ particles generated in the hadronic interactions, which decay into $\gamma$ rays that develop into EM showers. 
The hadronic component ($h$) consists of everything else such as protons, neutrons, and charged pions. 
The calorimeter response ($R_\mathrm{cal}$), defined as the ratio of the visible energy to the available energy, is very different between $e$ and $h$. As described in Sec.~\ref{sec:missing_energy}, missing energy due to nuclear breakup can be up to 30\%--50\% for hadrons with large event-by-event fluctuation, while for EM particles the response is typically higher with smaller fluctuations. 
A calorimeter is characterized by the $e/h$ ratio, defined as the ratio between $R_\mathrm{cal}^e$ and $R_\mathrm{cal}^h$. Calorimeters with $e/h=1$ are called ideal compensating calorimeters. The benefit of compensating calorimeters comes from the principle of calorimeter energy reconstruction:
\begin{equation} \label{eq:fem}
    E_\mathrm{rec} = \frac{f_{em} \cdot E_\mathrm{vis}}{R_\mathrm{cal}^e} + \frac{(1-f_{em}) \cdot E_\mathrm{vis}}{R_\mathrm{cal}^h} ,
\end{equation}
where $f_{em}$ is the fraction of the visible energy in the EM component and $E_\mathrm{rec}$ is the reconstructed energy of the incident particle.
Since $f_{em}$ has a large event-by-event fluctuation and is difficult to measure, it often dominates the calorimeter energy resolution. For ideal compensating calorimeters with $e/h=1$, however, Eq.~\ref{eq:fem} can be simplified to $E_\mathrm{rec} = E_\mathrm{vis} / R_\mathrm{cal}$, which does not suffer from $f_{em}$ fluctuations.

Compensation between the $e$ and $h$ responses is typically realized in sampling calorimeters using a combination of three techniques. First, the $e$ response can be reduced by using high-Z absorber materials. Second, the neutron response, which is a significant component in $h$, can be boosted by using hydrogenous active materials. Finally, the sampling fraction can be tuned to match $e$ and $h$ responses so that $e/h$ is close to 1. Examples of high-performance compensating colorimeters include the ZEUS calorimeter at HERA using uranium absorbers and plastic scintillators~\cite{ZEUSCalorimeterGroup:1989ill}, and the Spaghetti Calorimeter (SPACAL) at CERN using lead absorbers and scintillation fibers~\cite{Acosta:1991ap}.

\begin{figure}[htp]
  \centering
    \includegraphics[width=\columnwidth]{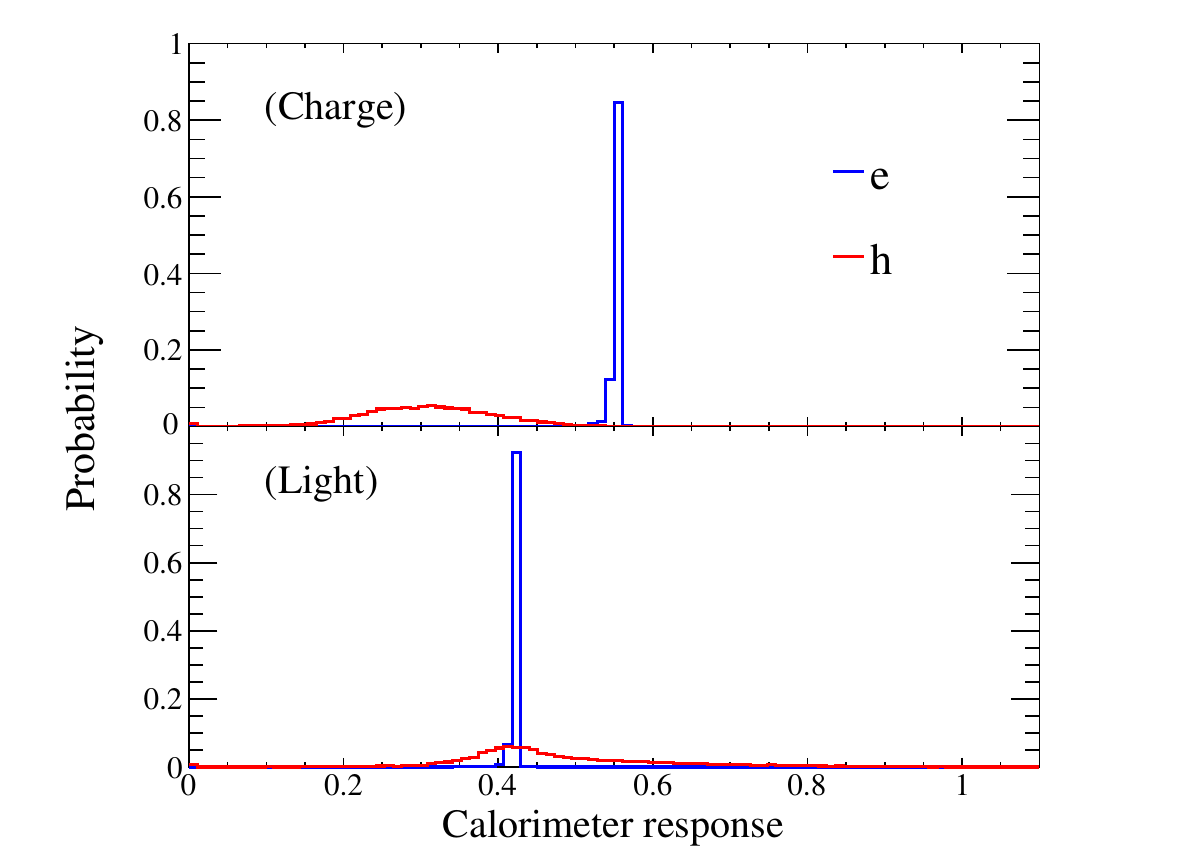}
  \caption{Calorimeter responses ($R_\mathrm{cal}$, the ratio of the visible energy to the available energy) in LArTPC at the drift E-field of 0.5~kV/cm for the EM component ($e$, blue) and the hadronic component ($h$, red). The top panel shows the charge calorimetry and the bottom panel shows the light calorimetry. $10^5$ $\nu_e$ CC events are simulated with a uniform neutrino energy distribution between 0.5--5~GeV. The area of each distribution is normalized to 1.
  }
  \label{fig:e-h}
\end{figure}

In a LArTPC neutrino detector, the $\nu_e$-Ar CC interaction (Eq.~\ref{eq:nueCC}) guarantees there are both EM (electrons and possibly $\pi^0$s) and hadronic components in the calorimeter. The fraction of the visible energy in the EM component ($f_{\mathrm{em}}$) varies event by event and is energy-dependent, which is modeled by the neutrino event generator such as GENIE. 
Similar to the problems of hadron calorimeters mentioned previously, if the LArTPC's calorimeter response ($R_\mathrm{cal}$) between $e$ and $h$ is very different, the fluctuation of $f_{\mathrm{em}}$ will largely impact the calorimeter energy resolution and the shape of the energy response. The calorimeter responses in LArTPC are studied with the detector simulation described in Sec.~\ref{sec:missing_energy}. For each $\nu_e$-Ar CC event, the electron and $\pi^0$s, if any, are grouped into $e$, and the protons, neutrons, pions, and other final-state particles are grouped into $h$. $R_\mathrm{cal}$ in each component is calculated and plotted in Fig.~\ref{fig:e-h}. 
The top panel shows the $e$ (blue) and $h$ (red) responses in the charge calorimetry and the bottom panel shows the light calorimetry. 
The event-by-event fluctuation in the $e$ responses is much smaller than the $h$ responses because of the much smaller contribution of missing energy in $e$. The most interesting feature, however, is that the peak of $e/h$ in the light calorimetry is close to 1, which makes it a compensating calorimeter. 
Since no extra material needs to be added to achieve compensation between the $e$ and $h$ responses, we call LArTPC a \emph{self-compensating} light calorimeter.

\begin{figure}[htp]
  \centering
    \includegraphics[width=\columnwidth]{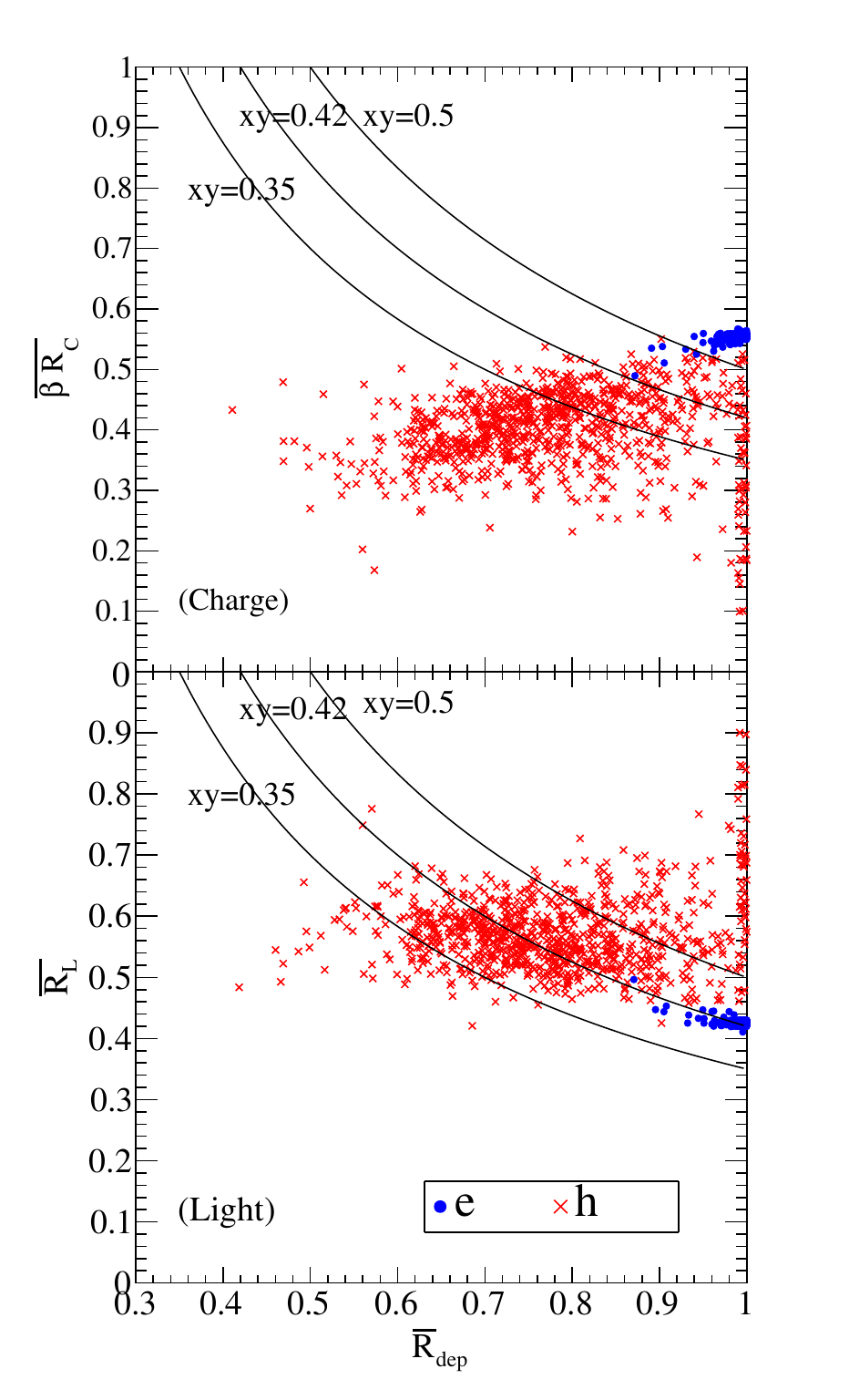}
  \caption{Scatter plot of 1000 simulated 3~GeV $\nu_e$-Ar CC events in a LArTPC at the drift E-field of 0.5~kV/cm. Activities in each event are grouped into the EM ($e$, solid circles) and hadronic ($h$, crosses) components. The y-axis shows the average recombination factor for charge ($\overline{\beta R_c}$, top) or light ($\overline{R_L}$, bottom) in $e$ (blue solid circles) or $h$ (red crosses). The x-axis shows the average ratio of the deposited energy to the total available energy in $e$ or $h$. The product of x and y is the calorimeter response shown in Fig.~\ref{fig:e-h}, where the $xy=0.42$ curve corresponds to the peak position of the $e$ or $h$ distribution in the light calorimeter response (bottom panel of Fig.~\ref{fig:e-h}).
  }
  \label{fig:RcRL_Rdep}
\end{figure}

The self-compensation mechanism in the scintillation light can be understood by rewriting the calorimeter response in two terms:
\begin{equation} \label{eq:R_cal}
    R_\mathrm{cal} = \frac{\sum E_\mathrm{vis}}{\sum \Edep} \cdot \frac{\sum \Edep}
    {\sum E_\mathrm{avail}} = 
    \begin{cases}
    \overline{\beta R_c} \cdot \bar R_\mathrm{dep}, \quad (\textrm{charge}) \\
    \overline{R_L} \cdot \bar R_\mathrm{dep}, \quad (\textrm{light})
    \end{cases}
\end{equation}
where the summation is over all particles in each component, $e$ or $h$. 
The first term in Eq.~\ref{eq:R_cal} is the average charge or light recombination factor, and the second term is the average ratio of the deposited energy to the total available energy. 
The two terms are plotted in Fig.~\ref{fig:RcRL_Rdep} for $e$ (blue solid circles) and $h$ (red crosses) separately from 1000 simulated 3~GeV $\nu_e$-Ar CC events, where the charge response is shown in the top panel and the light response is shown in bottom. The $\bar R_\mathrm{dep}$ is close to 1 for $e$ and $\sim$0.7 for $h$, showing that the missing energy fraction is significant in $h$. On the other hand, the light recombination factor ($R_L$), based on Eq.~\ref{eq:R_L} and Fig.~\ref{fig:recob}, is larger for $h$ than $e$ because of the higher $dE/dx$ for heavily ionizing particles such as protons in $h$. 
The extra scintillation light from recombination compensates for the larger missing energy in $h$, resulting in both the $e$ and $h$ responses centralized around the $R_\mathrm{cal}=xy=0.42$ curve in Fig.~\ref{fig:RcRL_Rdep}, making $e/h$ close to 1 in the light calorimetry. 
This effect, however, is anti-compensating for ionization electrons since the charge recombination factor $R_c$ is smaller for $h$ than $e$, making $e/h$ close to 1.8 in the charge calorimetry.

\begin{figure}[htp]
  \centering
    \includegraphics[width=\columnwidth]{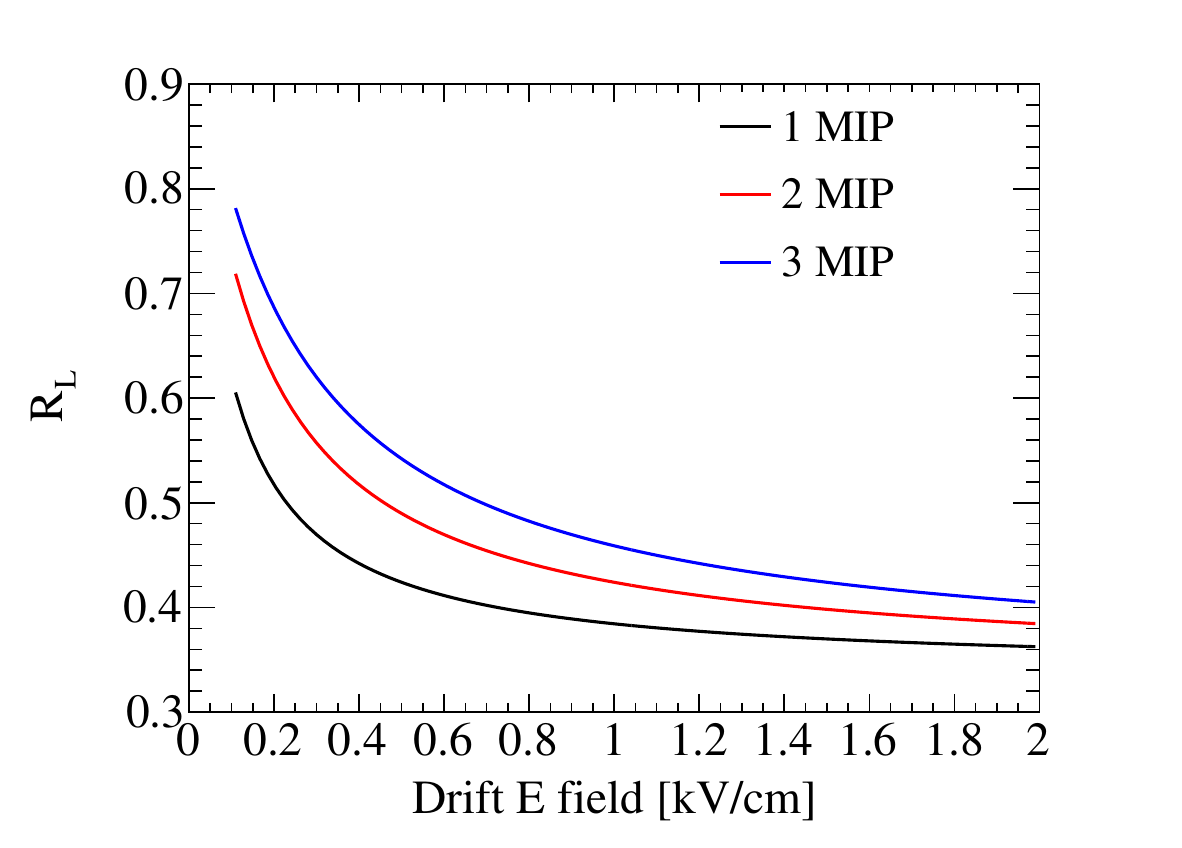}
    \includegraphics[width=\columnwidth]{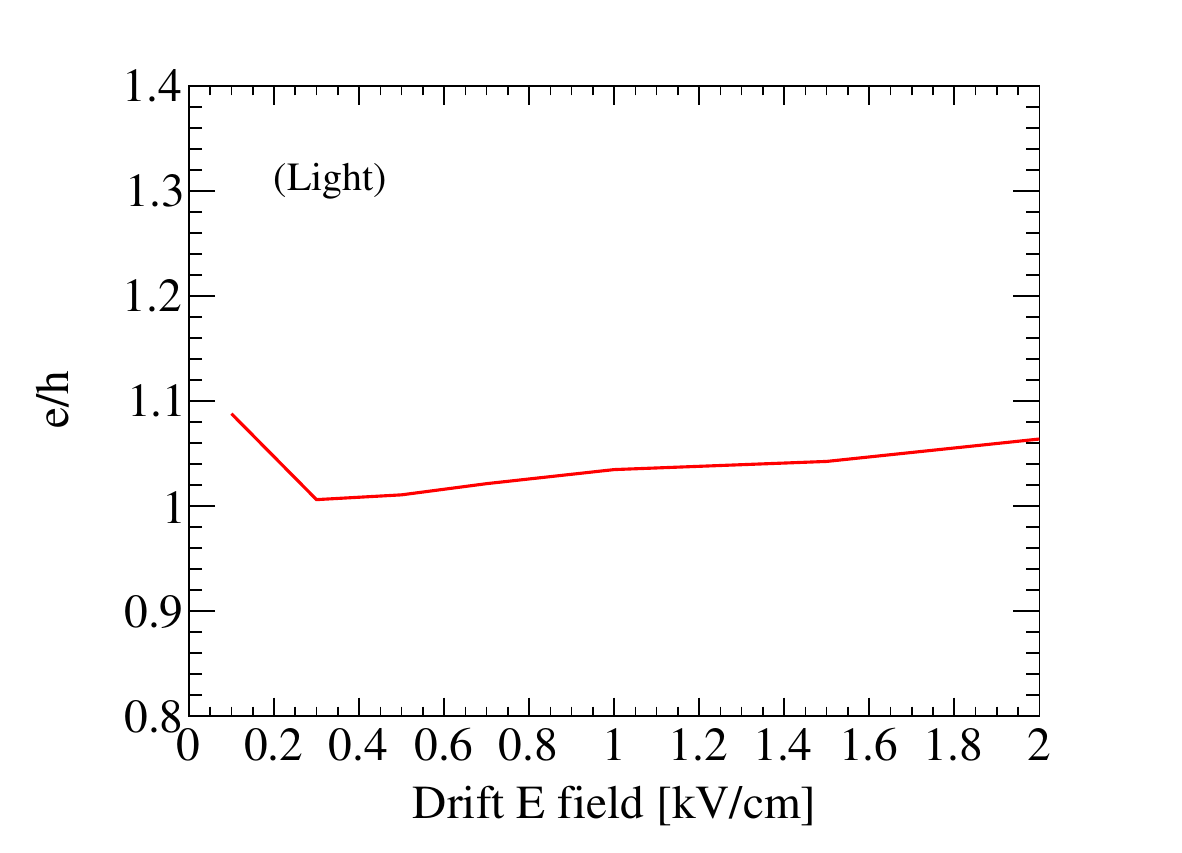}
  \caption{(Top) Light recombination factor ($R_L$) as a function of the drift electric field. The three curves correspond to particles with $dE/dx$ equal to one, two, and three times that of a minimum ionizing particle (2.1~MeV/cm), respectively. (Bottom) Simulated calorimeter response ratio between $e$ and $h$ ($e/h$) as a function of the drift electric field. For each drift electric field from 0.1--2~kV/cm (with 0.1 kV/cm increments), $10^4$ $\nu_e$-Ar CC events are simulated with a uniform neutrino energy distribution from 0.5--5~GeV.
  }
  \label{fig:E-field}
\end{figure}

The compensation performance of the LArTPC light calorimetry is dependent on the size of the drift electric field. The top panel of Fig.~\ref{fig:E-field} shows the light recombination factor ($R_L$) as a function of the drift electric field derived from Eqs.~\ref{eq:Rc} and \ref{eq:R_L}. 
The different colored curves show the $R_L$ for particles with $dE/dx$ equal to one, two, and three times that of a MIP ($dE/dx =  2.1$~MeV/cm). 
At a fixed drift electric field, the more separated are the $R_L$s, the more compensation there is between the $e$ and $h$ responses. 
To quantify the compensation performance, for each drift electric field from 0.1--2~kV/cm (with 0.1~kV/cm increments), $10^4$ $\nu_e$-Ar CC events are simulated in a LArTPC with a uniform neutrino energy distribution between 0.5--5~GeV. 
The $e/h$ is calculated for each E-field and plotted in the bottom panel of Fig.~\ref{fig:E-field}. The ideal compensation of $e/h$ close to 1 can be achieved from 0.3--0.5~kV/cm. The compensation is slightly worse at lower and higher E-fields, but good compensation of $e/h$ from 1--1.05 can still be achieved across a broad range of E-fields between 0.2 and 1.8 kV/cm. 

\begin{figure}[htpb]
  \centering
  \includegraphics[width=\columnwidth]{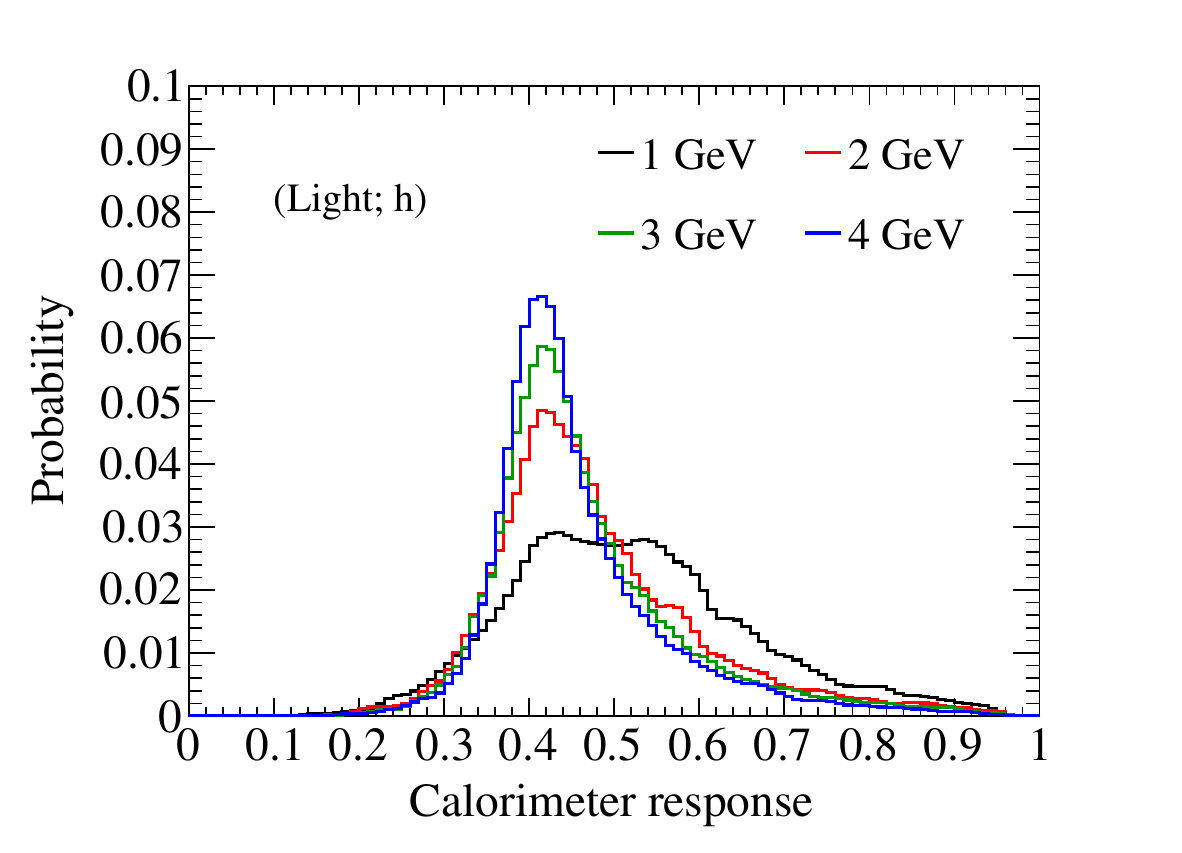}
  \caption{Simulated light calorimeter response of the hadronic component $h$ in LArTPC at the drift E-field of 0.5~kV/cm for different incident neutrino energies.
  }
  \label{fig:Rcal_E}
\end{figure}

The compensation performance of the LArTPC light calorimetry is also energy dependent. To study the energy dependence, the events in Fig.~\ref{fig:e-h} are divided into different incident neutrino energy bins, and the calorimeter response of the hadronic component ($R_\mathrm{cal}^h$) in the light calorimetry is plotted in Fig.~\ref{fig:Rcal_E}. 
For neutrino energies above 2~GeV, the $R_\mathrm{cal}^h$ distributions are similar, with smaller fluctuations for higher energy neutrinos. The $R_\mathrm{cal}^h$ distribution, however, is much wider for the 1~GeV or lower-energy incident neutrinos, which leads to a worse reconstructed energy resolution for lower-energy neutrinos (cf.~Fig.~\ref{fig:resolution}) using light calorimetry.

\begin{figure}[htpb]
  \centering
    \includegraphics[width=\columnwidth]{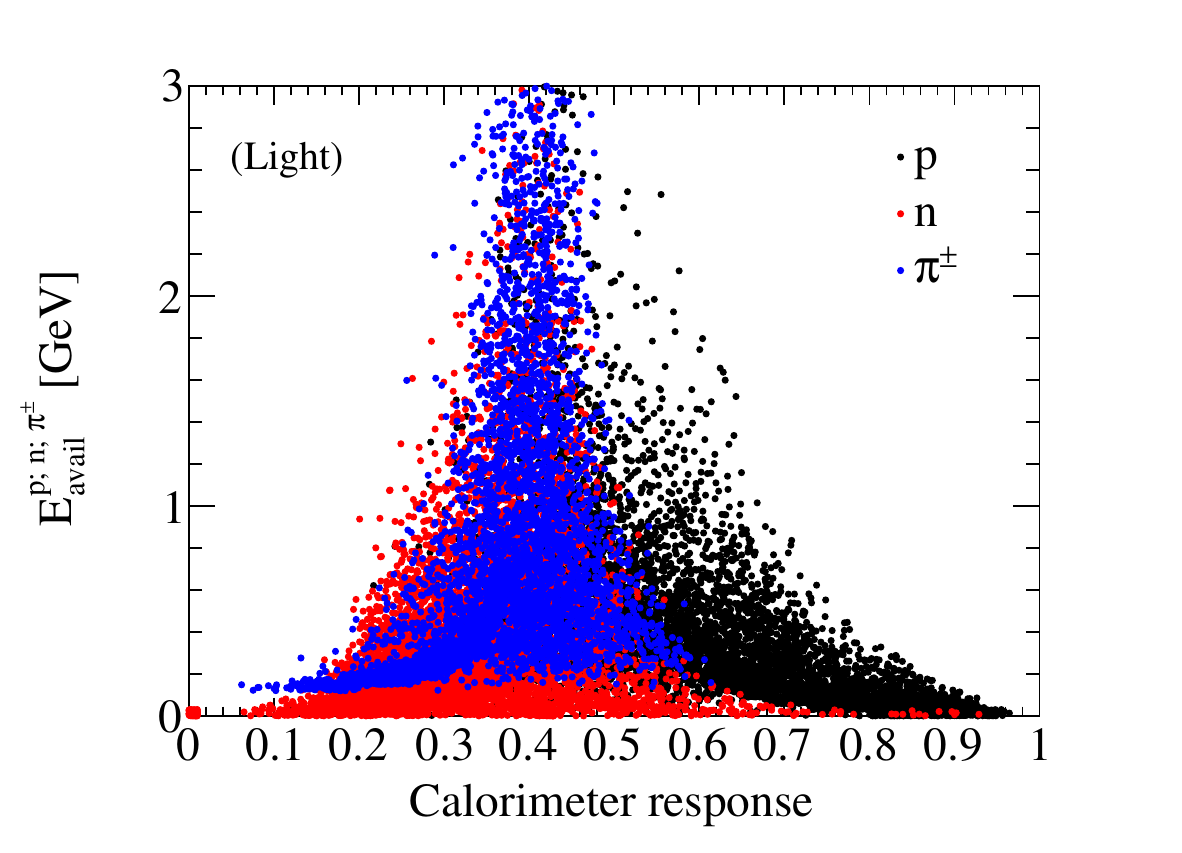}
    \includegraphics[width=\columnwidth]{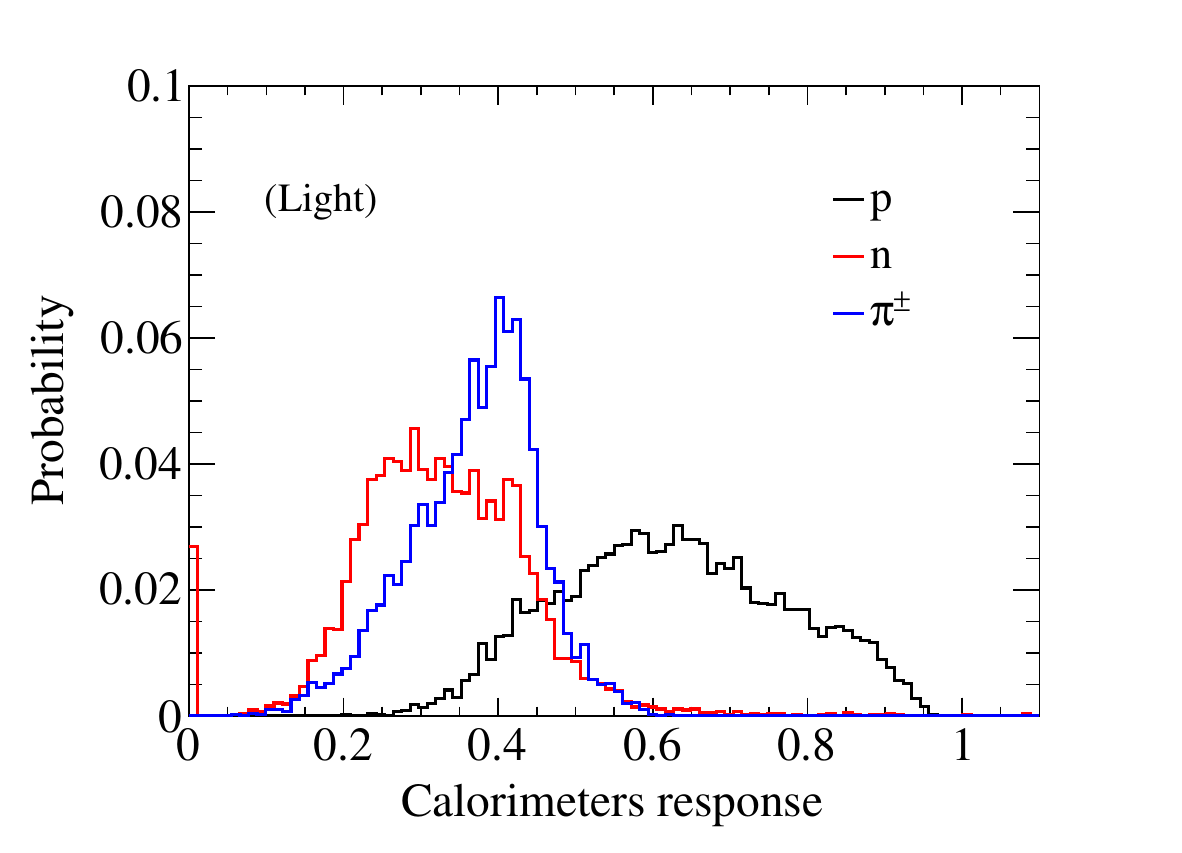}
  \caption{Simulated light calorimeter response ($R_\mathrm{cal}$) for protons, neutrons, and charged pions in LArTPC at the drift E-field of 0.5~kV/cm . The top panel shows a scatter plot of the available energy of each particle versus its $R_\mathrm{cal}$. The bottom panel shows the projected distribution of $R_\mathrm{cal}$ of different particles for all energies.
  }
  \label{fig:had_sep}
\end{figure}

This energy dependence is further studied in Fig.~\ref{fig:had_sep} by separately plotting the $R_\mathrm{cal}$ for protons, neutrons, and charged pions in the light calorimetry. The top panel shows a scatter plot of the available energy of each particle (i.e.~the kinetic energy of protons and neutrons, and the total energy of the charged pions) versus its $R_\mathrm{cal}$. The bottom panel shows the projected distribution of $R_\mathrm{cal}$ of different particles for all energies. 
We can observe that at low energy below $\sim$1~GeV, the behavior is quite different between $p$, $n$, and $\pi^\pm$. This is because a large contribution of $R_\mathrm{cal}$ comes from the primary particle itself before it possibly starts a hadronic shower. 
The proton has a higher $R_\mathrm{cal}$ because it starts as a charged particle without missing energy, and receives extra recombination light. 
The neutron has a lower $R_\mathrm{cal}$ because it has a high missing energy from the beginning. 
The charged pion has a similar $dE/dx$ to a MIP particle before it starts the hardronic shower, so its $R_\mathrm{cal}$ is similar to electrons, with additional missing energy from particle decays. Therefore, the total $R_\mathrm{cal}$ of the hadronic component becomes dependent on the distribution of final state particles, leading to a wider distribution for lower energy neutrinos as seen in Fig.~\ref{fig:Rcal_E}. However, at higher energy above $\sim$2~GeV, the $p$, $n$, and $\pi^\pm$ behave very similarly because they all start hadronic cascades and the $R_\mathrm{cal}$ is dominated by the secondary shower particles, which are similar for different primary particles. Thus, the compensation performance is better for higher-energy neutrinos and less sensitive to the modeling of final-state particles after neutrino interactions.

\section{Light calorimetry performance}
\label{sec:light-calo-performance}

Based on the discussions in Sec.~\ref{sec:L_calo} and Appendix~\ref{sec:app_charge_imaging}, in this section, we describe four energy reconstruction methods in LArTPC, one for light calorimetry and three for charge calorimetry, and compare their performance. The light calorimetry utilizes the self-compensating calorimeter concept described earlier.  The charge calorimetry methods are similar in concept to previous works from LArTPC experiments and are included here for comparison.

\begin{enumerate}
  \item \emph{L1}: Simple light calorimetry by summing up $L$ from all particle activities and scaling to the incident neutrino energy,
  \begin{equation} \label{eq:L1}
      \Erec^{L1} = \frac{L}{0.42}.
  \end{equation}
  
  \item \emph{Q1}: Simple charge calorimetry by summing up $Q$ from all particle activities and scaling to the incident neutrino energy,
  \begin{equation}
      \Erec^{Q1} = \frac{Q}{0.48}.
  \end{equation}

  \item \emph{Q2}: Assuming $e$ and $h$ charge activities can be grouped using pattern recognition algorithms on the 3D image (cf. Appendix~\ref{sec:app_charge_imaging}), scaling the $Q$ from $e$ and $h$ separately before summing up to the incident neutrino energy,
  \begin{equation} \label{eq:Q2}
      \Erec^{Q2} = \frac{Q_e}{0.55} + \frac{Q_h}{0.31}.
  \end{equation}
  
  \item \emph{Q3}: On top of Q2, assuming tracks longer than 2~cm can be reconstructed and the $dE/dx$ along the tracks can be measured, the $\Edep$ of the tracks can then be faithfully reconstructed by correcting for the charge recombination factors along the tracks. For the rest of the dot-like activities, scale the $Q$ from $e$ and $h$ separately similar to Q2:
  \begin{equation} \label{eq:Q3}
      \Erec^{Q3} = \Edep^{\mathrm{tracks}} + \frac{Q_e^\mathrm{dots}}{0.47} + \frac{Q_h^\mathrm{dots}}{0.20}.
  \end{equation}
\end{enumerate}

The scaling factors used in each reconstruction come from the detector simulation described in Sec.~\ref{sec:missing_energy}, which account for both the recombination effect and the missing energy contributions.

We also studied two additional reconstruction methods but they are not reported in detail here. The first method is the $Q+L$ calorimetry based on Eq.~\ref{eq:QL}. While it removes the recombination effect and recovers $\Edep$, it does not account for the missing energy contribution, therefore, its performance is only slightly better than Q1 but worse than Q2 and L1. 
The second method assumes activities from individual particles can be grouped and their energies can be reconstructed separately. 
This is mostly useful for muons and charged pions since their missing energies from decays or captures can be corrected event by event, using methods such as a range-based energy reconstruction. 
The missing energy fluctuation from nuclear breakup still remains, e.g.~from neutrons, making its performance only slightly better than Q3.

\subsection{Energy resolution comparision}
\label{sec:dual-calo-resolution}

\begin{figure}[htp]
  \centering
    \includegraphics[width=\columnwidth]{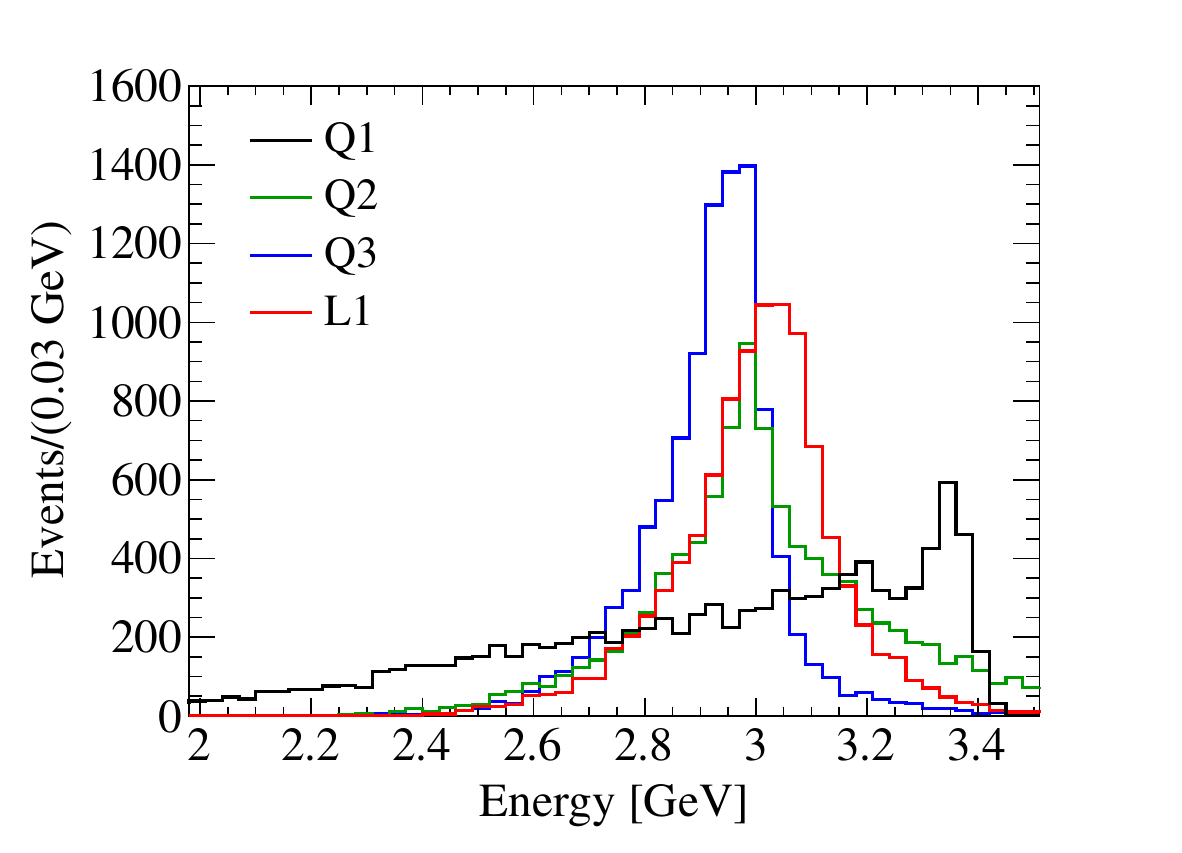}
    \includegraphics[width=\columnwidth]{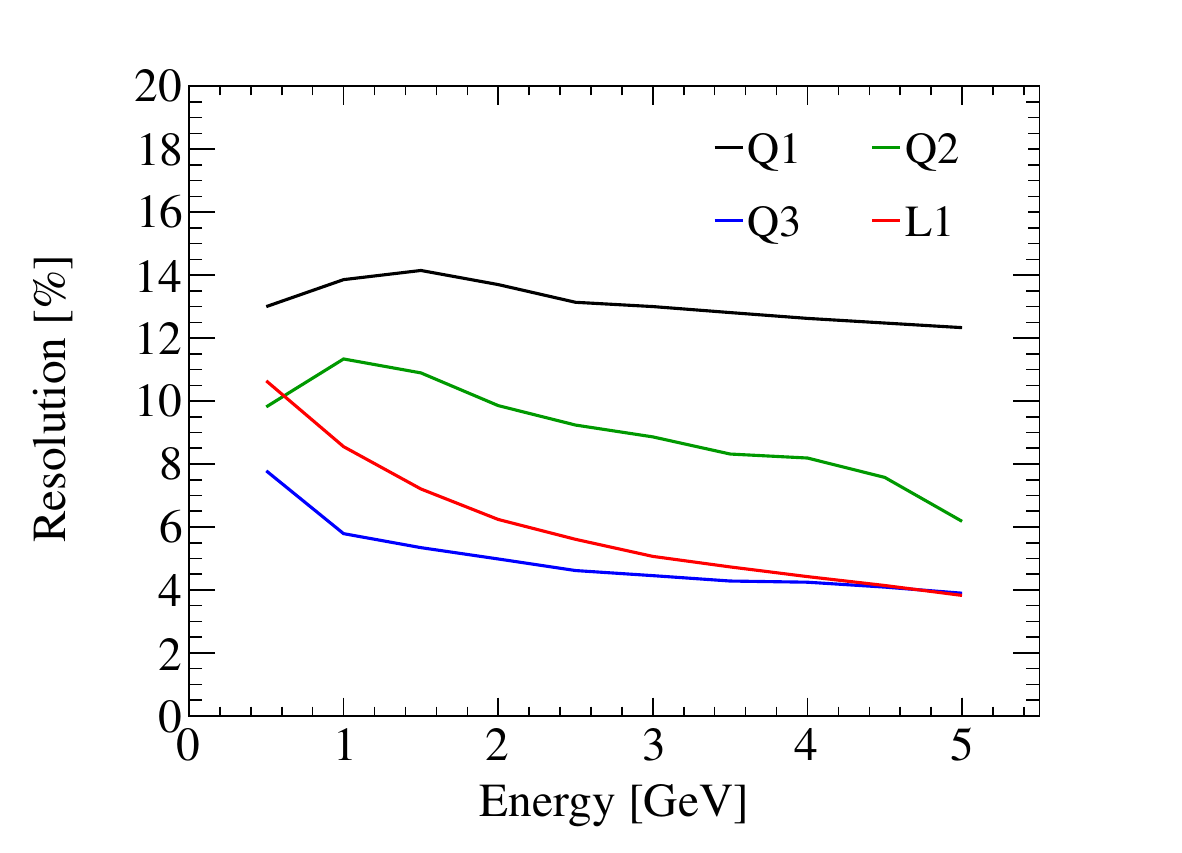}
  \caption{(Top) Distribution of the reconstructed neutrino energy from $10^4$ simulated 3~GeV $\nu_e$-Ar CC interaction events in LArTPC at the drift E-field of 0.5~kV/cm. The four colored histograms correspond to the four reconstruction methods Q1--3 and L1, respectively. (Bottom) Relative energy resolution, $\sigma/\bar E$, as a function of the incident neutrino energy for the four reconstruction methods Q1--3 and L1.
  }
  \label{fig:resolution}
\end{figure}

The top panel of Fig.~\ref{fig:resolution} shows the distribution of the reconstructed neutrino energy from $10^4$ simulated 3~GeV $\nu_e$-Ar CC interactions using the Q1--3 and L1 reconstruction methods. Q1 (black) has a long tail extending to low energies, which comes from the missing energy contribution and the different calorimeter response between $e$ and $h$. L1 (red) removes the low-energy tail through the self-compensation mechanism, resulting in a more symmetric shape and much better resolution. Q2 (green) has a similar performance to L1, sharper near the peak but with a longer tail in the high energy. Q3 (blue) has the best overall performance. Both Q2 and Q3 utilize LArTPC's charge imaging capability to separate $e$ and $h$ activities to improve the reconstructed energy resolution.

The bottom panel of Fig.~\ref{fig:resolution} shows the relative energy resolution, $\sigma/\bar E$, as a function of the incident neutrino energy. 
For each fixed neutrino energy, the $\sigma$ is calculated from the root mean square (RMS) of the distribution from each reconstruction method, and the mean energy of the distribution is used as $\bar E$. The RMS is larger than the width of the peak for a tailed distribution such as Q1 and Q2. This plot shows that Q3 achieves the best energy resolution at $\sim$6\% at 1~GeV and Q1 has the worst resolution of $\sim$14\% at 1~GeV. The L1 method has an 8.4\% resolution at 1~GeV. The Q2 method performs similarly to L1, although the longer tail at high energy makes its RMS worse than L1. 
In practice, the 2D energy response matrices of reconstructed neutrino energy vs.~true neutrino energy, also called the detector smearing matrices, are used in physics analysis, and they are shown in Fig.~\ref{fig:response_matrix} in Appendix~\ref{sec:app_response_matrix}. The energy resolution and response matrices from this study can be readily compared with other literature or experimental works~\cite{DeRomeri:2016qwo,Friedland:2018vry,DUNE:2020ypp,MicroBooNE:2024zhz,Kopp:2024lch}, and can be used in the phenomenology studies such as those in Ref.~\cite{DeRomeri:2016qwo,DUNE:2021cuw}. One should keep in mind the experimental conditions assumed in different studies.

It's worth noting that although Q3 has the best overall performance, the light calorimetry L1 is a simpler reconstruction method without needing complex pattern recognition algorithms to separate $e$ versus $h$, or track-like versus dot-like activities. For higher energy neutrinos above 3~GeV, the performance between L1 and the much more sophisticated Q3 methods is nearly equal owing to the self-compensation mechanism in the light calorimetry. 
Therefore, the light calorimetry in LArTPC provides an independent and comparable energy estimation, complementing the well-established charge calorimetry. 
In addition, since the light calorimetry through the self-compensation mechanism has very different systematic uncertainties compared to the charge-based calorimetry, such as its dependence on the final state particle distribution (cf.~Fig.~\ref{fig:had_sep}) from neutrino event generator models~\cite{Andreopoulos:2009rq}, it provides a new way of estimating and mitigating the systematic uncertainty in GeV neutrino energy reconstruction, which is one of the largest uncertainties affecting the measurements of neutrino oscillation parameters and $\nu$-Ar cross sections~\cite{MicroBooNE:2024kwe}.

\subsection{Impact of nonuniform light collection} \label{sec:app_nonuniformity}

In the previous studies, the light yield is assumed to be uniform across the entire detector volume, which in practice is difficult to achieve because of the constraints of photon detector placement and achievable surface coverage. Variation of light collection efficiency in large detectors presents a well-known challenge in scintillation-based energy reconstructions. Here we study the impact of nonuniform light collection on the energy resolution in LArTPC light calorimetry.

The nonuniformity of the photon collection efficiency (PCE) depends on specific detector designs. As an example, Fig.~15 of Ref.~\cite{DUNE:2024wvj} shows a GEANT4 simulation of the APEX design for DUNE Phase-II far detectors. 
The average PCE in the APEX detector is 0.8\%, leading to an average light yield of $\sim$180 photoelectrons per MeV. 
The photon detector locations are shown in Fig.~12 of Ref.~\cite{DUNE:2024wvj}. While the light yield is rather uniform across the majority of the detector volume, it is smallest in the middle near the cathode and on the top and bottom near the anodes ($\sim$109 PE/MeV), and largest near the photon detectors at the field cage walls of the LArTPC ($\sim$300 PE/MeV). 
Given this variation of light collection, a simple light calorimetry reconstruction as described by Eq.~\ref{eq:L1} would not yield good results without position-dependent event selections, and additional steps are necessary to correct for the light collection nonuniformity. 
We note that efforts are ongoing to optimize the design of the APEX detector including methods such as adding photon detectors at the cathode and increasing the reflectivity of the anodes, all of which would further improve the uniformity of the light collection.

In an ideal event reconstruction, given a known photon collection efficiency map $\epsilon(x,y,z)$, the light nonuniformity can be corrected by calculating a correction factor $C$ for each event to obtain the effective PCE of this event ($\epsilon_{\mathrm{eff}} = C \cdot \bar\epsilon$),
\begin{equation}  \label{eq:C}
    C = \frac{\sum_i (dL)_i \epsilon_i}{\bar\epsilon L},
\end{equation}
where $i$ goes through each voxel position of the reconstructed energy deposition. In the numerator, $(dL)_i$ is the reconstructed visible light in voxel $i$ and $\epsilon_i$ is the PCE in voxel $i$. In the denominator, $L$ is the total visible light and $\bar\epsilon$ is the average PCE of all voxels. While $\epsilon_i$ can be obtained from detector simulation and a well-designed calibration program, the reconstruction of $(dL)_i$ is known to be difficult unless for simple event topologies such as point-like events (low-energy electrons), or line-like events (high-energy muons). 
For more complex event topologies such as the neutrino interaction shown in Fig.~\ref{fig:lartpc-2D}, the reconstruction of $(dL)_i$ at a fine position resolution is challenging using light signals alone. 

One of the main advantages of the LArTPC is its superb 3D imaging capability with mm-scale position resolution using ionization charge signals. This charge-based event reconstruction provides a new way to perform light nonuniformity correction, which is not available in other scintillator detectors that only record light signals. 
For each event, we can calculate a different correction factor $C'$ using charge-based reconstruction:
\begin{equation} \label{eq:C'}
    C' = \frac{\sum_i (dQ)_i \epsilon_i}{\bar\epsilon Q},
\end{equation}
where $(dL)_i$ is replaced by $(dQ)_i$ --- the reconstructed visible charge in each voxel $i$.
Correspondingly, the total visible light $L$ is replaced by the total visible charge $Q$. Since $(dQ)_i$ can be well reconstructed with high position resolution~\cite{Qian:2018qbv,MicroBooNE:2020vry,MicroBooNE:2017xvs,MicroBooNE:2016dpb}, $C'$ can be calculated to a much better precision than $C$. If $(dQ)_i/(dL)_i$ is a constant, then $C'$ would equal $C$ and the light nonuniformity can be perfectly corrected. However, since $(dQ)_i$ and $(dL)_i$ have a different dependence on $dE/dx$ (cf.~Fig.~\ref{fig:recob}), there is residual nonunformity after using Eq.~\ref{eq:C'}, which needs to be quantified.

\begin{figure}[htp]
  \centering
    \includegraphics[width=\columnwidth]{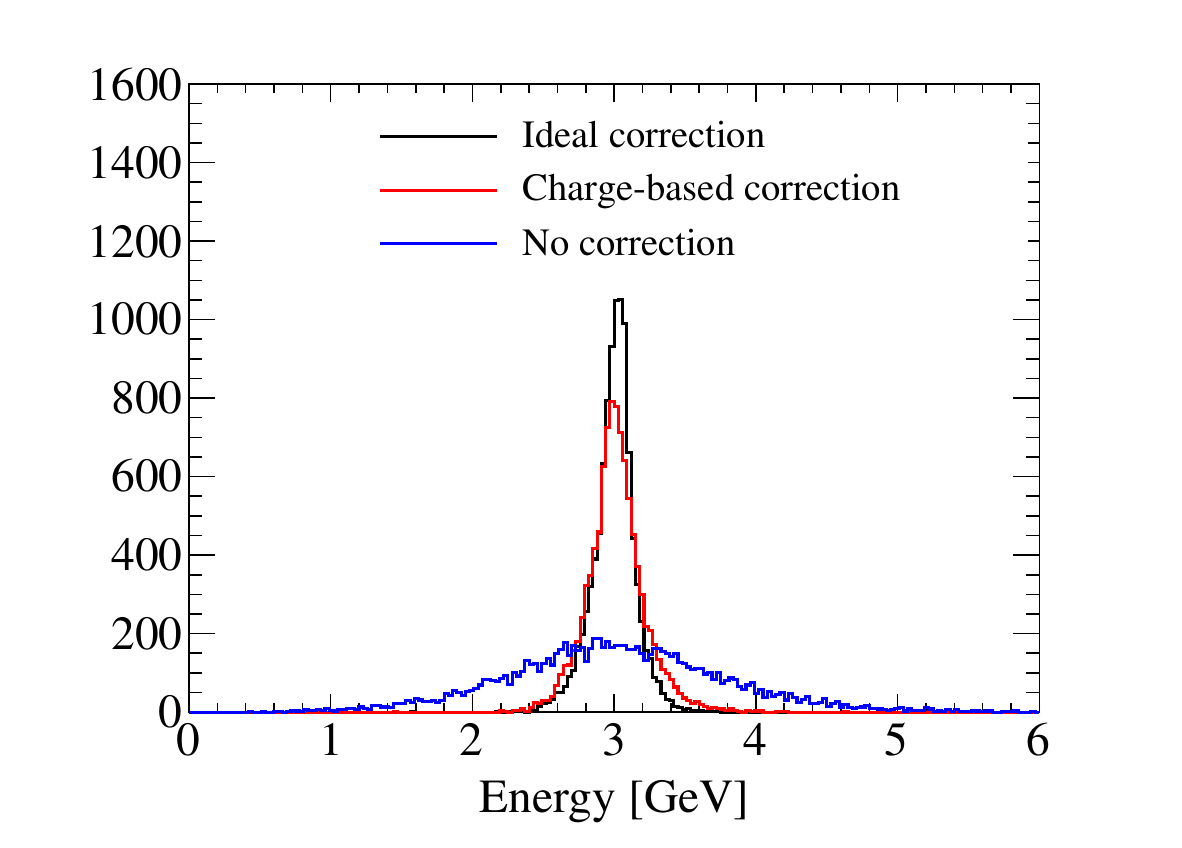}
    \includegraphics[width=\columnwidth]{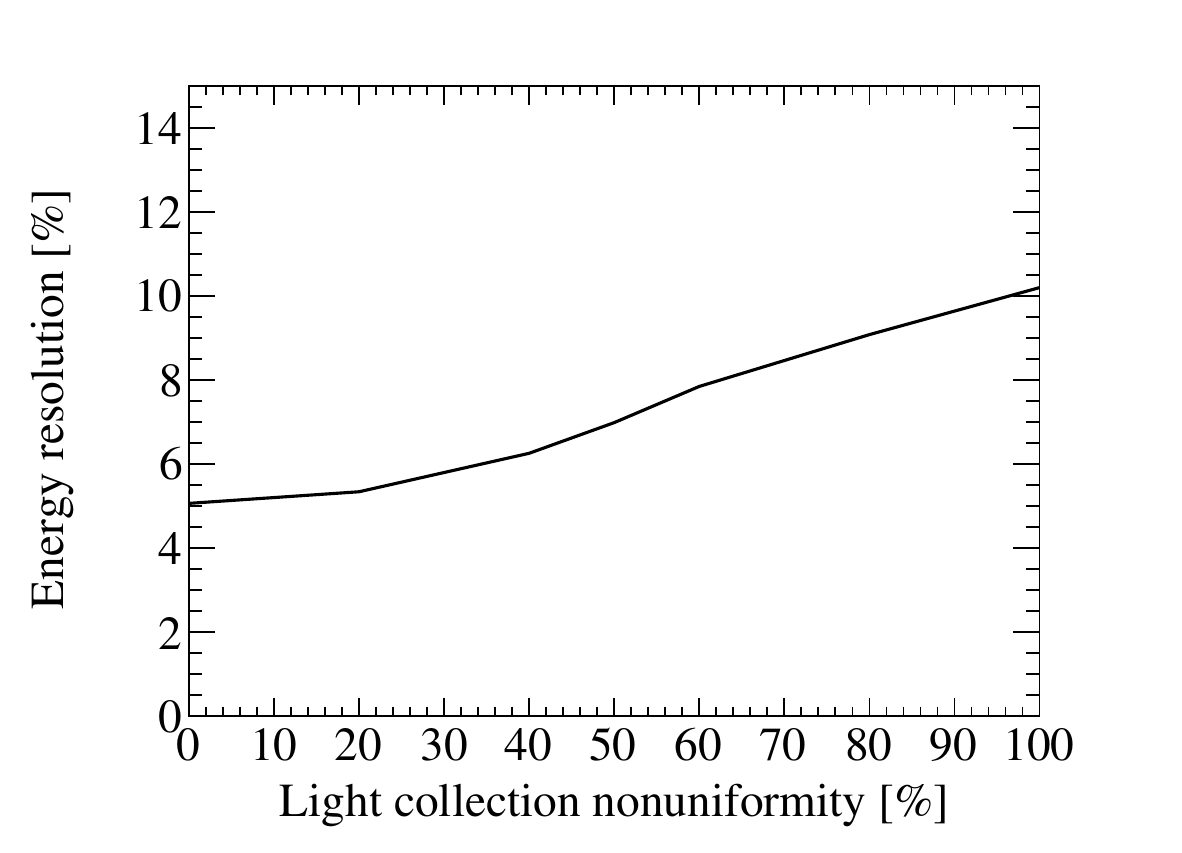}
  \caption{
    (Top) Reconstructed neutrino energy distributions using light calorimetry from $10^4$ simulated 3 GeV $\nu_e$-Ar CC interactions with 50\% variation of position-dependent photon collection efficiency. 
    The blue histogram assumes an average PCE following Eq.~\ref{eq:L1} without any nonuniformity corrections. 
    The black histogram uses an ideal nonuniformity correction following Eq.~\ref{eq:C}. 
    The red histogram uses the charge-based nonuniformity correction following Eq.~\ref{eq:C'}.
    (Bottom) Reconstructed neutrino energy resolution of 3~GeV $\nu_e$-Ar CC interactions from light calorimetry using charge-based nonuniformity correction versus the size of the nonuniformity.
  }
  \label{fig:nonunformity}
\end{figure}

To quantify the impact of nonuniform light collection and be more general, we divided the detector volume into 10-cm bins along one dimension $x$. In each bin, a photon collection efficiency $\epsilon(x)$ is randomly generated following a Gaussian distribution with a mean of 0.8\% and 1$\sigma$ fluctuation from 0\% up to 100\%. The top panel of Fig.~\ref{fig:nonunformity} shows the results for 3~GeV $\nu_e$-Ar CC interactions from a 50\% position-dependent PCE variation. 
The blue histogram shows the reconstructed neutrino energy distribution using an average PCE following Eq.~\ref{eq:L1} without any nonuniformity corrections. 
The nonuniformity variation dominates energy resolution in this case. The black histogram shows an ideal correction using Eq.~\ref{eq:C}. 
The reconstructed energy distribution returns to the case of uniform light collection. 
The red histogram shows the distribution after charge-based correction using Eq.~\ref{eq:C'}. We can see that the nonuniformity effect is largely removed, and the energy resolution is only worsened from 5.1\% to 7.0\% even at a 50\% PCE variation. The bottom panel of Fig.~\ref{fig:nonunformity} further shows the reconstructed neutrino energy resolution of 3~GeV $\nu_e$-Ar CC interactions using charge-based nonuniformity correction versus the size of the nonuniformity, demonstrating the effectiveness of charge-based correction method. 
This study can be generalized to a 3D PCE map $\epsilon(x,y,z)$ with a similar conclusion. In reality, the variation of PCE is smooth in nearby voxels, and the local variation of PCE is not as drastic as in our simulation. Therefore, our studies here represent a worst-case scenario and demonstrate that the charge-based nonuniformity correction provides a practical solution for the light calorimetry in LArTPC. 
We also note that this correction (Eq.~\ref{eq:C'}) does not require particle identification level information (such as $dE/dx$ along a trajectory, $e$ versus $h$ separation, etc.), therefore the implementation is relatively straightforward and less sophisticated than the charge imaging calorimetry reconstruction methods such as Q2 and Q3 described by Eqs.~\ref{eq:Q2} and \ref{eq:Q3}, which require pattern recognition and other techniques to distinguish different particle activities.

\section{Discussions and outlook}
\label{sec:conclusions}

In this work, we have reviewed the LArTPC's capability as a dual calorimeter to estimate particle energy from both ionization charge and scintillation light. 
We demonstrate that because of the recombination luminescence, the LArTPC functions as a self-compensating light calorimeter: the missing energy in the hadronic component is compensated for by the increased luminescence relative to the electromagnetic component. 
Using 0.5--5~GeV electron neutrino charged current interactions as a case study, we show that good compensation performance of $e/h$ from 1--1.05 can be achieved across a broad range of drift electric fields (0.2--1.8~kV/cm), with better performance for neutrino energies above 2~GeV.
Under ideal conditions of uniform light collection, we show that LArTPC light calorimetry can achieve an energy resolution comparable to the charge imaging calorimetry. Challenges arising from nonuniform light collection in large LArTPCs can be mitigated with a position-dependent light yield correction derived from 3D charge signal imaging.


We made a few simplifications in this study. The LArTPC is assumed to be infinitely large so that all particle activities (except neutrinos) are contained. For finite-sized detectors, the missing energy from partially contained particles should be studied. 
The LArTPC is assumed to be fully active, which is not the case for detectors with multiple drift volumes. 
The active volume for charge and light calorimetry is assumed to be the same, but scintillation light can also be generated in LAr outside of the TPC active volume. 
The scintillation light is assumed to be fully collected regardless of its time distribution, but the late light component could play an important role in the photon signal reconstruction and affect the calorimeter response~\cite{DUNE:2024dge}. 
Other contributions to energy resolution such as electronic noise, constant backgrounds, and channel-to-channel variations, are also ignored in this work. While the event-by-event fluctuation in the missing energy is often the dominating factor in the energy resolution, these other contributors could still play a large role if the detector design is not optimized for energy resolution. 
Therefore, the results from this work can be viewed as an optimal case for the performance of light calorimetry in LArTPC. Dedicated simulation and analysis studies are necessary to understand all contributions to energy resolution for specific detector designs. 

The results from this work can have multiple applications:
\begin{enumerate}[leftmargin=*]
    \item As a theoretical framework to properly implement scintillation light simulation in LArTPC experiments. While comprehensive models, such as NEST~\cite{Szydagis:2011tk}, to simulate the scintillation yield of noble liquids are popular in dark matter experiments, the light simulation in most of the current LArTPC neutrino experiments is less sophisticated without considering recombination luminescence. This work demonstrates why a proper light simulation is important in LArTPC light calorimetry and outlines the detailed implementation procedures. 

    \item As a motivation to perform light calorimetry in LArTPC experiments for GeV neutrino physics. In most of the current LArTPC neutrino experiments, scintillation light is only used to provide the $t_0$ and trigger information. Light calorimetry to improve MeV neutrino physics has been suggested~\cite{DUNE:2020ypp}, but the motivation to conduct high-energy light calorimetry is lacking. This work demonstrates that in a LArTPC experiment with a capable light detection system, light calorimetry can provide an independent energy measurement with a comparable performance to the more sophisticated charge imaging calorimetry for GeV neutrinos. This will benefit numerous physics programs, such as long-baseline neutrino oscillations, atmospheric neutrinos, tau neutrino physics, and many others.


    \item As a new approach to evaluating and mitigating the systematic uncertainties in neutrino energy measurements with LArTPCs. The systematic uncertainty associated with neutrino energy reconstruction is one of the largest uncertainties affecting the measurements of neutrino oscillation parameters and $\nu$-Ar cross sections. Two main contributors to the uncertainty in neutrino energy reconstruction are 1) mistakes in the pattern recognition algorithms when analyzing the charge-based images, and 2) dependence on the final state particle distribution provided by $\nu$-Ar interaction event generator models. The simplistic nature of the light calorimetry and less dependence on the final state distribution (cf.~Sec.~\ref{sec:L_calo}) provides a new way to tackle this problem.
    
    \item As a foundational model to be tested. The self-compensation mechanism described in this work can be readily tested in the current and future LArTPC experiments with capable photon detection systems, such as ICARUS~\cite{ICARUS-T600:2020ajz}, SBND~\cite{SBND:2020scp}, the DUNE LAr near detector modules~\cite{DUNE:2024fjn}, and the DUNE far detector prototypes~\cite{DUNE:2021hwx,DUNE:2024wvj}. Care should be taken to calibrate the photon detector response and correct for effects such as the light collection nonuniformity, time integration dependence, and the selection of events.

    \item As a design direction to improve light detection capability in future LArTPC experiments. Optimized photon detector designs, such as the APEX detector concept, are being carried out in the DUNE Phase-II program~\cite{DUNE:2024wvj} for the next far detectors to improve their light detection capability such as optical coverage, photon collection efficiency, uniformity of the light response, and position resolution, among others. While the original purpose for such an upgrade was to improve DUNE's lower energy physics program such as nucleon decay, supernova neutrino detection, and solar neutrinos, the results from this work provide another strong motivation and reference on how to optimize the photon detector system design to further enhance DUNE's capability in long-baseline neutrino oscillation and other high-energy physics programs. 
    
    \item As a starting point for works beyond neutrino physics. While this work focuses on the energy reconstruction of few-GeV neutrinos, it can be extended to the calorimetry of higher energy particles since the principles presented are general. The self-compensation mechanism in the LArTPC light calorimetry provides a new way to build compensating calorimeters without reducing the performance of EM responses. Coupled with the sampling calorimeter technology to contain the particle activities in a small LArTPC, scintillation light calorimetry could be considered on its own or as complementary to the well-established ionization charge calorimetry as the next-generation calorimeter for high energy collider experiments. Dedicated simulation studies are necessary to optimize its design to reach the best achievable energy resolution.
    
\end{enumerate}

While LArTPC has been traditionally considered only as a charge-based calorimeter for GeV neutrino physics, we hope this work has proven its potential as a scintillation light calorimeter and serves as a stepping stone for future applications of light calorimetry with LArTPCs. 


\bigskip

\begin{acknowledgments}
We thank Xin Qian and Milind Diwan for providing valuable discussions on this research. This work is supported by the U.S.~Department of Energy (DOE) Office of Science, Office of High Energy Physics under Contracts No.~DE-SC0012704, No.~DE-SC0009854; BNL LDRD No.~23-058; and the 2024 SBU-BNL Seed Grant Program.
\end{acknowledgments}

\appendix

\section{Quanta reducing processes}
\label{sec:app_quanta_reducing}

There are three main processes that could reduce the total number of quanta ($N_e + N_{ph}$) in a LArTPC. All of them can be neglected in the scope of this work:

\paragraph{Escaping electrons} At zero electric field, the photon yield was observed to be only about 40000 photons per MeV, i.e.~$\sim$78\% of the ideal photon yield~\cite{Doke:1990rza,Marinho:2022xqk}. This effect was attributed to the ``escaping electrons'' --- some electrons are thermalized outside of the ion-influence radius and can live for a long time ($>$ milliseconds)  without recombination, thus escaping the light collection time window. This effect diminishes once a drift electric field is applied, where the escaping electrons are collected as charge signals. 

\paragraph{Scintillation quenching} It was observed that for $\alpha$ particles, Au ions, and fission fragments at high LET, the total quanta ($N_e + N_{ph}$) collected is only about 20\%--70\% of the ideal yield. 
This was explained by the biexicitonic quenching model~\cite{Hitachi:1992iyr}, where two $\Ar^*$ atoms can combine and de-excite nonradiatively, i.e.~dissipate their energy through heat. 
The biexicitonic quenching factor not only depends on LET, but also on the track microscopic structure such as the radial distribution of the energy deposition. 
While this is an important systematic effect for nuclear recoils in low energy dark matter experiments, the biexicitonic quenching was calculated to be negligible for relativistic particles and for few-MeV protons~\cite{Hitachi:1992iyr}.

\begin{figure}[htp]
  \centering
    \includegraphics[width=\columnwidth]{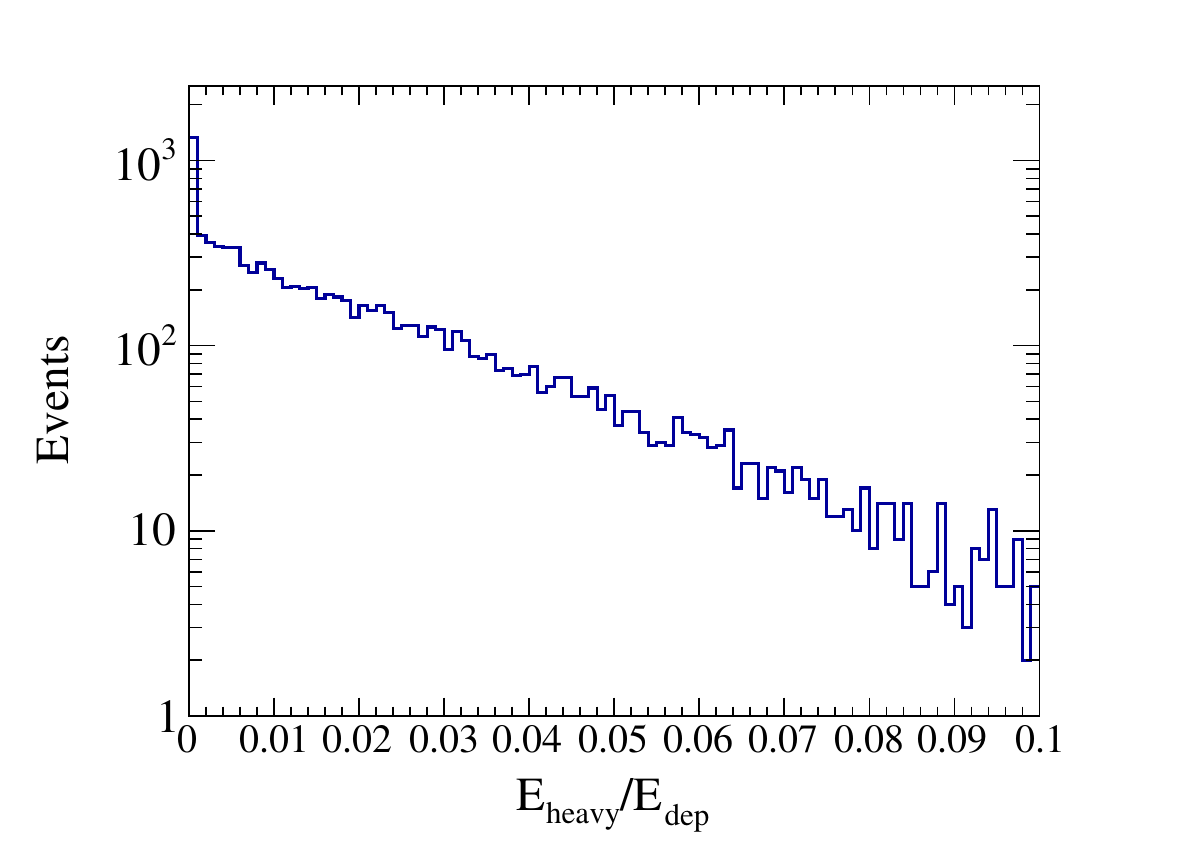}
    \includegraphics[width=\columnwidth]{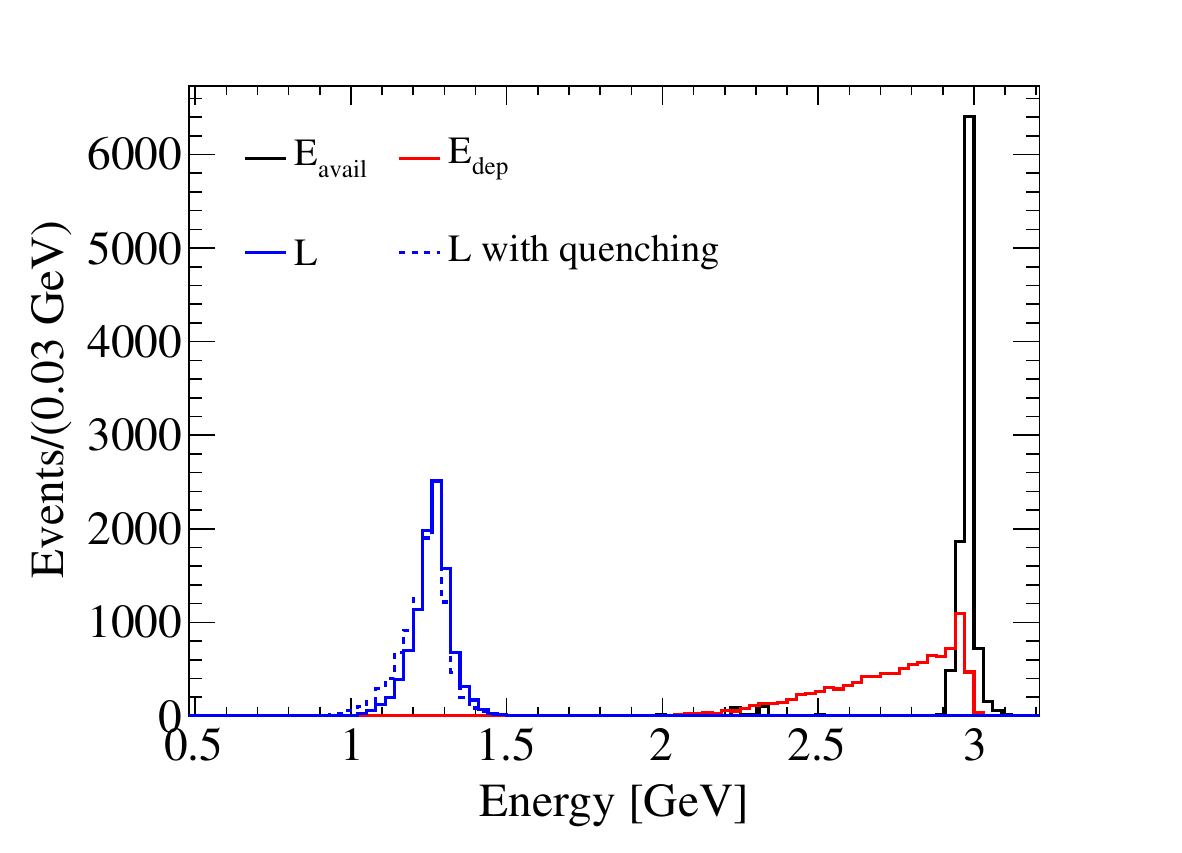}
  \caption{
  (Top) Distribution of the energy deposited by heavy particles ($E_\mathrm{heavy}$), such as $\alpha$ particles and recoiled nuclei, divided by the total deposited energy ($\Edep$) from $10^4$ simulated 3~GeV $\nu_e$-Ar CC interactions.   
  (Bottom) The dashed blue curve shows the visible energy in light (L) after applying quenching factors for heavy particles. The other three curves are the same as in Fig.~\ref{fig:edep_QL} and are included for comparison.
  }
  \label{fig:nr}
\end{figure}

In GeV $\nu$-Ar interactions, heavy particles such as $\alpha$ particles or recoiled ions (Ar, Si, Cl, etc.) may be released following a nuclear breakup or secondary nuclear scattering, with kinetic energies typically below a few MeV. Figure \ref{fig:nr} shows the impact of the scintillation quenching from these heavy particles. The top panel shows the distribution of the energy deposited by heavy particles ($E_\mathrm{heavy}$) divided by the total deposited energy ($\Edep$) from $10^4$ simulated 3~GeV $\nu_e$-Ar CC interactions. The distribution peaks near zero, indicating that heavy particle contribution is small in GeV neutrino interactions. To study the scintillation quenching impact, we applied a quenching factor (QF) of 0.7 to $\alpha$ particles~\cite{Doke:2002oab}. For recoiled heavy ions, the QF is energy dependent. At 10--100 keV recoiled energy, the QF is between 0.25 and 0.35~\cite{Segreto:2020qks}. For MeV recoiled energy, experimental data are less abundant, and we applied a QF of 0.25 as the maximum amount of quenching and 0.7 as the minimum amount of quenching for MeV heavy ions to study their impact. In the bottom panel, the light visible energy ($L$) distribution after applying the maximum amount of quenching is shown as the dashed blue histogram. Compared to the solid blue histogram without considering quenching, the distribution is slightly broadened in the low-energy tail. 
The energy resolution of the light calorimetry for 3 GeV neutrino is worsened from 5.1\% to 6.1\% after applying the maximum amount of quenching for heavy particles (cf.~Fig.~\ref{fig:resolution}). If we use a QF of 0.7 for MeV heavy ions instead of 0.25, the energy resolution is only worsened to 5.4\%. 
Since the impact on energy resolution is small ($<1$\%), we have neglected the scintillation quenching in the studies presented in the main text and added the discussion here as a systematic uncertainty to consider. 

A second type of scintillation quenching happens when parts-per-million level nitrogen contamination exists in LAr. In this case, the $\Ar_2^*$ excimer states could de-excite by collision with N$_2$ molecules~\cite{WArP:2008rgv}, thus reducing the scintillation light yield. We assume the nitrogen impurity can be well controlled and do not consider this type of scintillation quenching in this work.

\paragraph{Electron attachment} Ionization electrons could attach to electronegative impurities in LAr and form negative ions during their passage to the anodes. Since the drift velocity of ions is about five orders of magnitude lower than that of electrons, these ions escape the charge collection and the original signal is attenuated as a function of drift distance. 
Common electronegative impurities in LAr are the oxygen and water molecules. The electron attachment rates for different impurities are summarized in Ref.~\cite{Li:2022pfu}. 
For large LArTPCs, purification of LAr to remove O$_2$ and H$_2$O is essential to allow a long drift distance of a few meters without significant signal attenuation. 
Parts-per-trillion level impurity and electron lifetime of more than 3~ms have been successfully demonstrated in recent experiments such as MicroBooNE~\cite{MicroBooNE:2016pwy} and ProtoDUNE-SP~\cite{DUNE:2021hwx}, so in this work we assume the electron attenuation from attachment to impurities is negligible.  

\section{Charge imaging calorimetry}
\label{sec:app_charge_imaging}

In Sec.~\ref{sec:L_calo}, we explain that in calorimeters where the $e$ and $h$ responses are very different, the event-by-event fluctuation in $f_{em}$ could largely impact the resolution of the reconstructed energy. Compensating calorimeters, such as the LArTPC light calorimetry described therein, mitigate this issue by making $e/h$ close to 1. An alternative approach is to measure the $e$ and $h$ components separately, and apply the calorimeter response corrections accordingly: 
\begin{equation} \label{eq:dual-readout}
    E_\mathrm{rec} = \frac{E_\mathrm{vis}^e}{R_\mathrm{cal}^e} + \frac{E_\mathrm{vis}^h}{R_\mathrm{cal}^h} ,
\end{equation}
where $E_\mathrm{rec}$ is the reconstructed energy of the incident particles. In the past HEP accelerator experiments, one way to achieve this is using dual-readout calorimeters that measure both the scintillation and the Cerenkov light~\cite{Akchurin:2005an,Lee:2017shn}. Since most of the deposited energy in a hadronic shower is by nonrelativistic particles such as low-energy protons, the Cerenkov calorimeter is less responsive to $h$. Therefore, by measuring both the scintillation and the Cerenkov light, the $f_{em}$ can be deduced event by event, and the $e$ and $h$'s calorimeter responses can be applied separately. 

\begin{figure}[htp]
  \centering
    \includegraphics[width=\columnwidth]{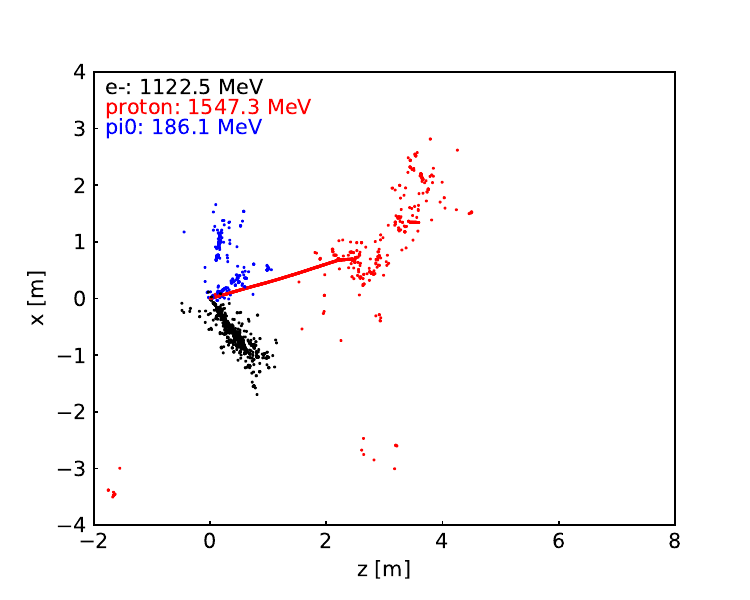}
  \caption{Example 2D event display of a simulated 3~GeV $\nu_e$ CC interaction in LArTPC. Different colors represent the activities from different primary particles generated by the neutrino interaction. 
  }
  \label{fig:lartpc-2D}
\end{figure}

In a LArTPC, the $e$ and $h$ activities in ionization electrons can be naturally separated utilizing LArTPC's superb 3D imaging capability with mm-scale position resolution. 
Figure \ref{fig:lartpc-2D} shows an example 2D event display of a simulated $\nu_e$-Ar CC interaction in LArTPC. 
Colors represent activities from different primary particles generated by the neutrino interaction. In this example, an electron, a $\pi^0$, and a proton are generated after the interaction, and the particles further develop into EM and hadronic showers in the detector.  
With dedicated LArTPC pattern recognition algorithms such as Wire-Cell~\cite{MicroBooNE:2020vry}, Pandora~\cite{MicroBooNE:2017xvs}, and Deep Learning~\cite{MicroBooNE:2016dpb}, among others, the $e$ (e.g.~the electron activities in black and the $\pi^0$ activities in blue) and $h$ (e.g.~the proton activities in red) components can be grouped separately, and Eq.~\ref{eq:dual-readout} can be used to reconstruct the neutrino energy. This method, also called the \emph{charge imaging calorimetry}, is adopted by most of the current LArTPC experiments for energy reconstruction, which can achieve better performance than the simple charge calorimetry by adding up all charges. 

The charge imaging calorimetry can be further improved by correcting for the $dE/dx$ dependent recombination factor $R_c$ event-by-event, where the fluctuation, in particular in $h$, could be large, as illustrated in Fig.~\ref{fig:RcRL_Rdep}. The $dE/dx$ can be calculated by reconstructing all track-like objects and measuring the $dE$ for each segment of $dx$ along the reconstructed trajectories. Then, the deposited energy for track-like objects can be faithfully reconstructed using Eq.~\ref{eq:E_dep}. However, if the track is too short, it becomes a dot-like object and the $dE/dx$ cannot be well measured since the $dx$ is difficult to reconstruct. Short tracks are commonly seen in the EM and hadronic showers that create many low-energy particles. 
The capability to reconstruct short tracks depends on the position resolution of the LArTPC, which typically has a pitch size of 3--5~mm and a track of a few centimeters can be well reconstructed. 
The separation of track-like and dot-like objects has been achieved by dedicated pattern recognition algorithms, in particular with the recent advancement in deep learning neural networks~\cite{MicroBooNE:2020yze,DUNE:2022fiy}.


\section{Detector smearing matrix} \label{sec:app_response_matrix}

\begin{figure}[htpb]
  \centering
    \includegraphics[width=0.48 \columnwidth]{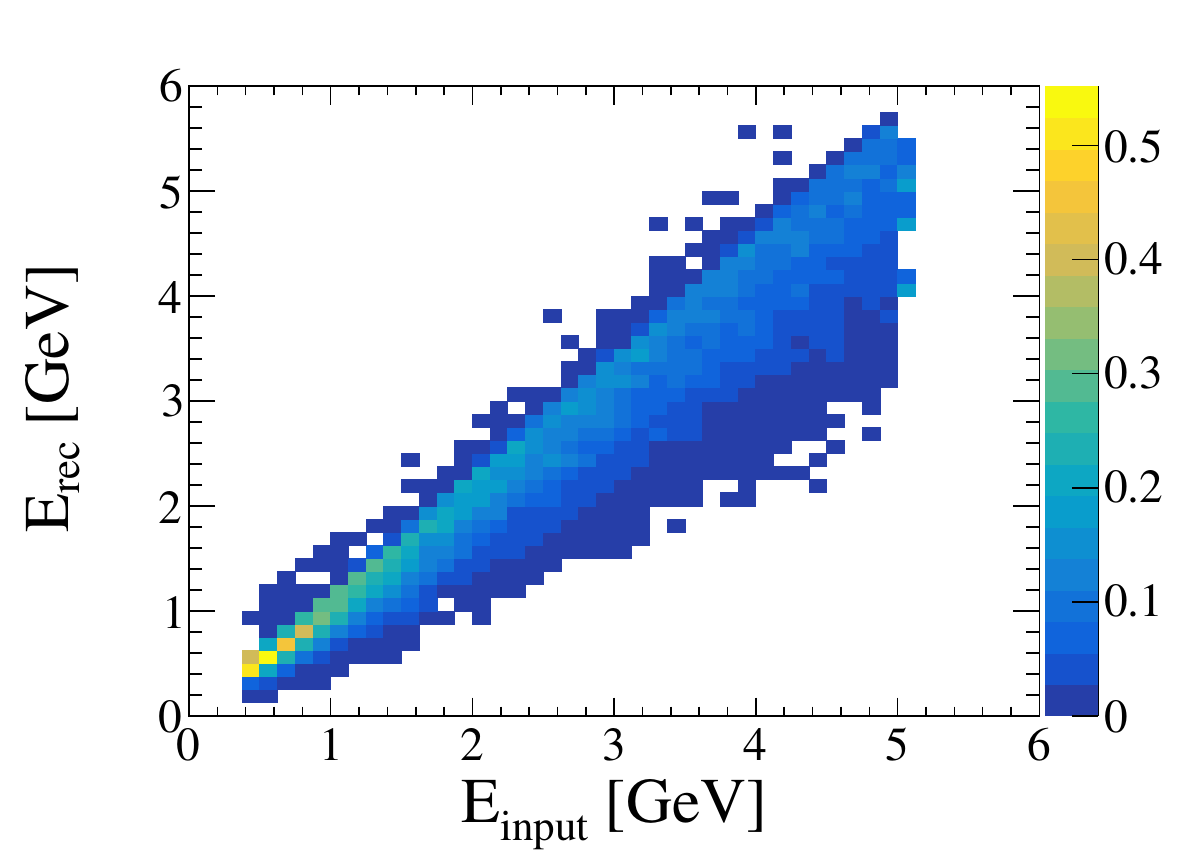}
    \includegraphics[width=0.48\columnwidth]{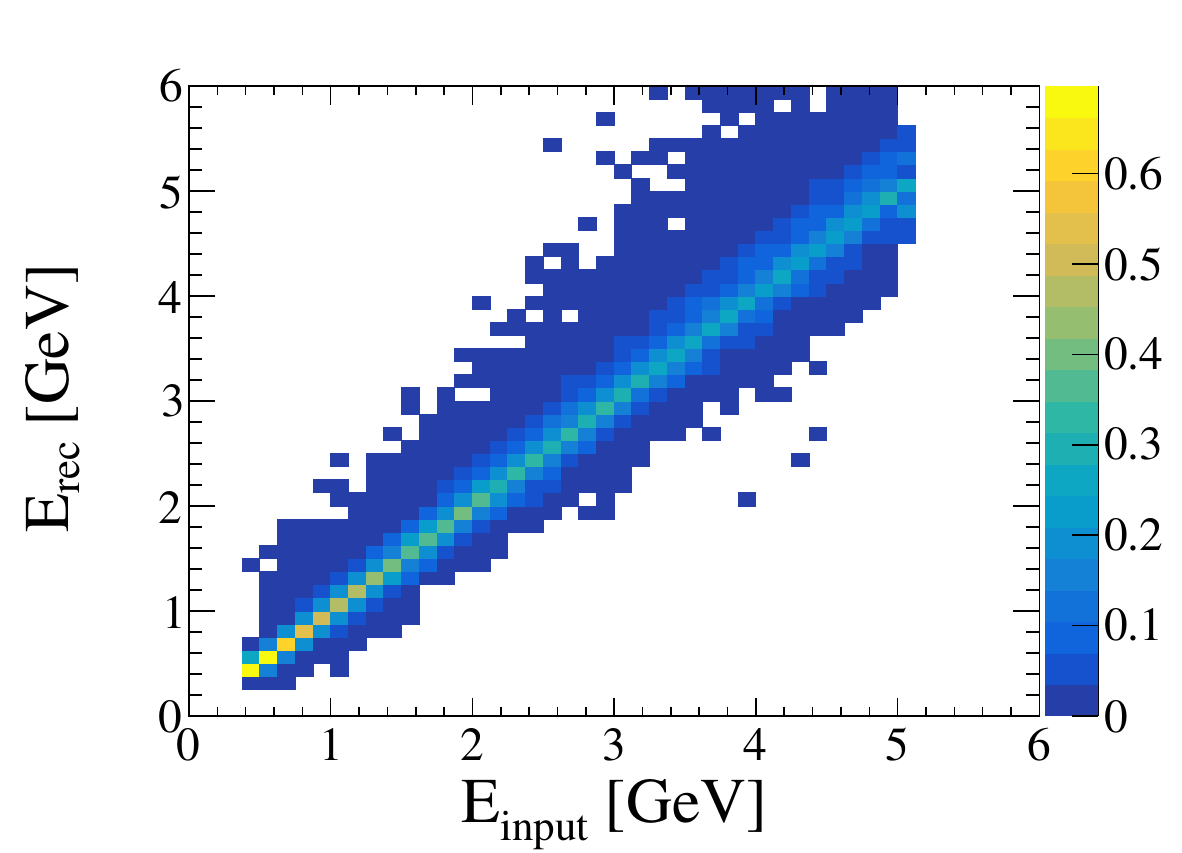}
    \includegraphics[width=0.48\columnwidth]{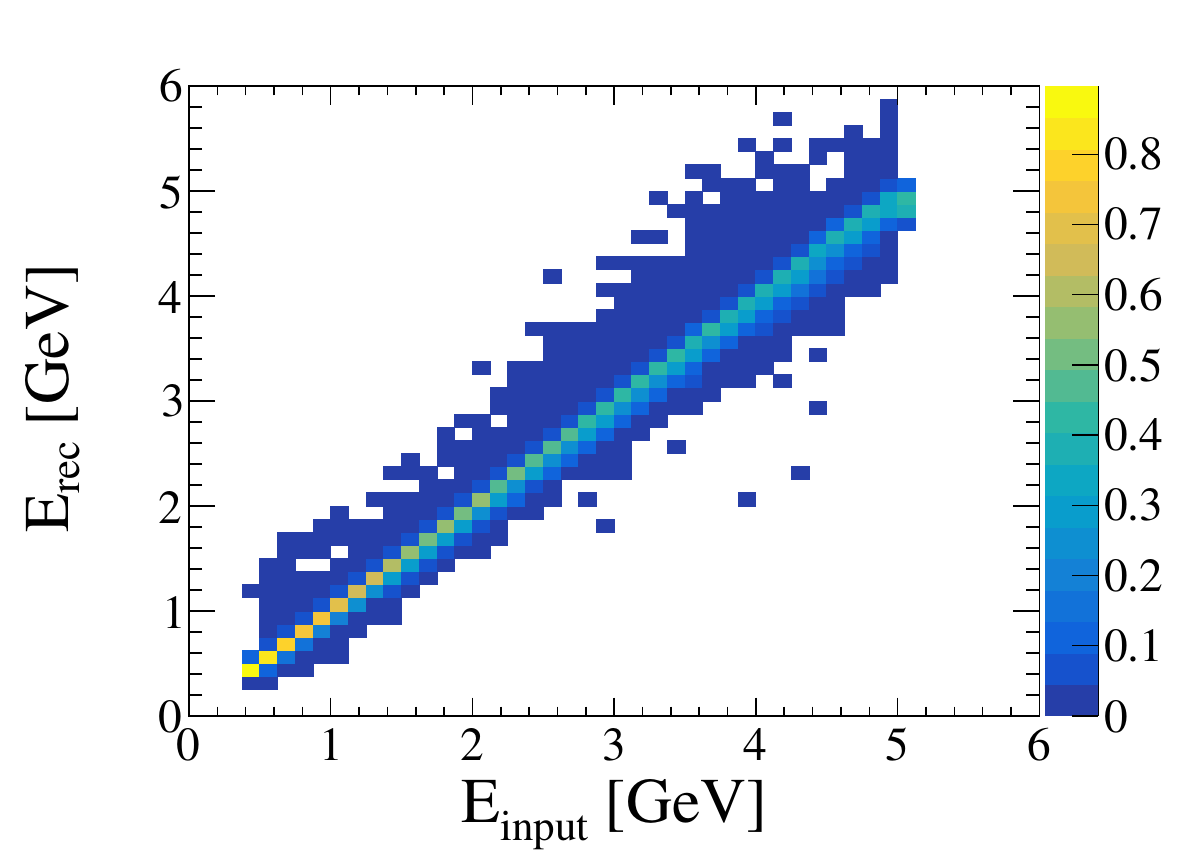}
    \includegraphics[width=0.48\columnwidth]{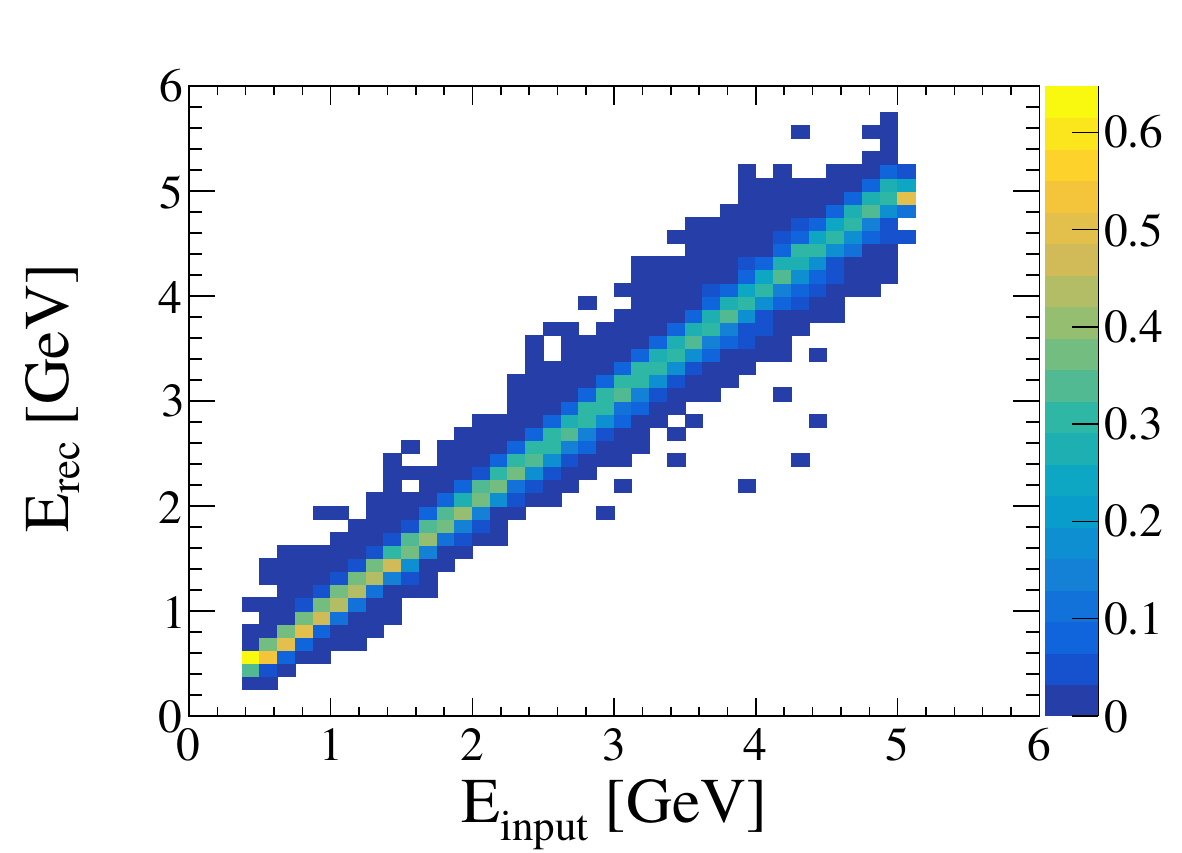}
  \caption{Detector smearing matrices of reconstructed neutrino energy vs.~true neutrino energy for electron neutrinos. The four subplots are for the four reconstruction methods Q1 (top left), Q2 (top right), Q3 (bottom left), and L1 (bottom right), respectively. 
  }
  \label{fig:response_matrix}
\end{figure}

Figure~\ref{fig:response_matrix} shows the detector smearing matrices of reconstructed neutrino energy vs.~true neutrino energy for $\nu_e$s. The four subplots are for the four reconstruction methods Q1--3 and L1, respectively. The smearing matrices for $\bar\nu_e$s are similar and not shown here. 

\bibliography{ref}

\end{document}